\pdfoutput=1

\documentclass[10pt]{article}   	% use "amsart" instead of "article" for AMSLaTeX format
\usepackage{geometry}                		% See geometry.pdf to learn the layout options. There are lots.
\usepackage[pdftex]{graphicx}
\usepackage{cite}
\usepackage{hyperref}
\usepackage{breakurl}
\usepackage{amsmath,amssymb,amsfonts}
\usepackage{algorithmic}
\usepackage{textcomp}
\usepackage{multirow} 
\makeatletter
\let\MYcaption\@makecaption
\makeatother
\usepackage[font=footnotesize]{subcaption}
\usepackage{bm}
\usepackage{authblk}
\providecommand{\keywords}[1]{\textbf{\textit{Index terms---}} #1}
\title{Structural Vulnerability in Y00 Protocols}
\author{Kentaro IMAFUKU}
\affil[]{National Institute of Advanced Industrial Science and Technology (AIST)}

\date{\empty}							% Activate to display a given date or no date

\begin{document}
\maketitle
\begin{abstract}
This paper critically analyzes the Y00 protocol, a quantum noise-based stream cipher proposed to enhance classical cryptographic methods through quantum mechanical properties. Despite its promise, we reveal a structural vulnerability that enables the leakage of secret information from measurement outcomes. To systematically evaluate its security, we first formalize the claims of previously proposed Y00 protocols, clarifying their achievements and limitations. We then identify the structural vulnerability through an intuitive explanation and rigorous formulation using maximum likelihood estimation. Our findings demonstrate that Y00's structural weaknesses allow for the unique determination of the shared secret, leading to significant information leakage. Using the "Toy protocol" as a reference model, we contextualize these results within the broader field of security technology. Furthermore, we generalize our findings to a wider class of quantum-based stream cipher protocols, identifying a fundamental security condition that Y00 fails to satisfy. This condition serves as a critical benchmark for ensuring the security of any stream cipher protocol relying on physical states, whether quantum or classical. These findings underscore the importance of rigorous security evaluations, particularly in systems intended for practical applications. Unexamined vulnerabilities not only undermine trust but also expose systems to avoidable risks, making rigorous analysis indispensable for ensuring resilience and security.
\end{abstract}
%\section{}
%\subsection{}

\section{Introduction}
\label{sec:introduction}
The advancement of science and technology is sustained by a continuous cycle of proposal and verification. Presenting new ideas is a remarkable human endeavor, worthy of respect on its own. However, science and technology cannot progress through proposals alone; it is only through rigorous verification by numerous individuals, including the proposers themselves, that the true value of a proposal is established.

If a novel idea is not appropriately evaluated, three scenarios are possible: A. it delivers the expected results, B. it has no substantial effect, or C. its design has unintended consequences that hinder the intended goals. Particularly problematic is scenario C—when an approach is mistakenly assessed as A under ambiguous reasoning— as this can harm the sound relationship between science, technology, and society. In today's highly information-driven society, the negative impacts of misjudging technologies related to information security can be severe and, in some cases, catastrophic. Consequently, it is essential that new ideas undergo careful verification, grounded in scientific rigor rather than mere perception or appearance.

This paper reexamines the security of the Quantum Noise Stream Cipher (QNSC), also known as the Y00 protocol. The Y00 protocol has inspired extensive research and development efforts, highlighting the potential of quantum-based cryptographic systems. However, rigorous evaluation reveals that its security claims require reconsideration. This paper focuses on the protocols defined in Secs. \ref{sec:Y00 introduction} and \ref{sec: choice of state}, which formalize the QNSC. These protocols define the use of shared secret keys and basis-dependent quantum states to encode transmitted data. For instance, they describe how ciphertexts are encoded into quantum states using basis-dependent mappings, as detailed in Secs. \ref{sec:Y00 introduction} and \ref{sec: choice of state}. To the best of the authors' knowledge, this formalization encompasses all variations and implementations previously referred to as QNSC or Y00, spanning a wide range of studies and proposals \cite{barbosa03,barbosa03_2,corndorf05,banwell05,hirota05,kato05,nair06,yuen07,nair08,shimizu08,kanter09,nakazawa14,yoshida15,yoshida16,jiao17,futami19,tanizawa19,futami20,tan20,iwakoshi20,yoshida21,chen21,iwakoshi21,sohma22,zhang22,hanwen24}. Despite claims in the literature of achieving security enhancements through quantum properties, our analysis identifies significant structural vulnerabilities in its design.

Generally, the security of secure communication is classified into two categories: “information-theoretic security” and “computational security.” The former considers an attacker with unlimited computational resources, while the latter relies on limitations in the attacker's computational capabilities. In traditional symmetric-key cryptography, Shannon's theorem establishes that transmitting more information than the entropy of the pre-shared key cannot achieve information-theoretic security \cite{shannon49}. When the Y00 protocol was initially proposed, it was suggested that by harnessing quantum properties, it could enable secure communication surpassing Shannon's limit, an appealing claim at the time \cite{barbosa03,barbosa03_2,corndorf05,banwell05,hirota05,kato05,nair06}. However, this claim was soon refuted \cite{nishioka04,yuan05,nishioka05,hoi05,tregubov21}. 
Furthermore, consistent with the findings of this paper, it has been preliminarily demonstrated that the protocol can be compromised without the use of known plaintexts \cite{donnet06,ahn07,ahn08}, highlighting that the security of Y00 critically depends on employing a strong PRNG \cite{yuen07,nair08,shimizu08}.

More recently, however, some proponents have argued that Y00 achieves a new level of security through a combination of computational security and physical layer security, leveraging the quantum states used as the transmission medium\cite{kanter09,nakazawa14,yoshida15,yoshida16,jiao17,futami19}. This approach is based on a concept of "dual structure," where a standard symmetric-key cipher produces ciphertext, which is then transmitted via quantum states. Based on this concept, active research and development efforts have been vigorously pursued  \cite{tanizawa19,futami20,tan20,iwakoshi20,yoshida21,chen21,iwakoshi21}.
In Japan, this concept has received positive assessments in government reports \cite{tezuka23}, and large-scale projects related to Y00's implementation have been initiated \cite{nedo24}. These developments suggest an acceleration towards Y00's practical application.

However, while claims regarding the enhanced security and resilience of the Y00 protocol may seem plausible, they lack rigorous formalization and a clear framework for discussing their security properties. This ambiguity hinders the iterative cycle of proposal and verification essential for healthy cryptographic development. To address this, we formalized the security claims of the previously proposed Y00 protocols and critically analyzed their achievements and limitations. Through this process, we identified a fundamental structural vulnerability in Y00, which we present in this paper.

Although there are several variations of the Y00 protocol, each uses quantum states determined by the transmitted data and the basis. The security of Y00 relies on shared basis information between legitimate communicators, rendering these states indistinguishable to attackers lacking the basis information. According to its proponents, this security is rooted in the principles of quantum mechanics, theoretically providing resistance even to attackers with quantum computing capabilities \cite{sohma22}. This paper's findings challenge this assertion, highlighting the following points:

\begin{enumerate} 
    \item Indistinguishability in Y00 relies not on quantum mechanical principles but rather on the entropy of pre-shared information used to share basis data, showing no connection to resistance against quantum computing.
    \item Y00 exhibits a structural flaw that allows for the leakage of pre-shared secret information under very weak conditions, even without known plaintext.
    \item As a result, the entropy contributing to secrecy in Y00 is reduced by the entropy expended to share basis information, making Y00 inferior to standard stream ciphers that use the same pre-shared secret.
\end{enumerate}
These findings, discussed in detail in Sec. 4, underscore the structural weaknesses of Y00 and its inability to achieve the security enhancements claimed by its proponents.

\begin{figure}[ht]
\begin{center}
\includegraphics[width=0.9\linewidth]{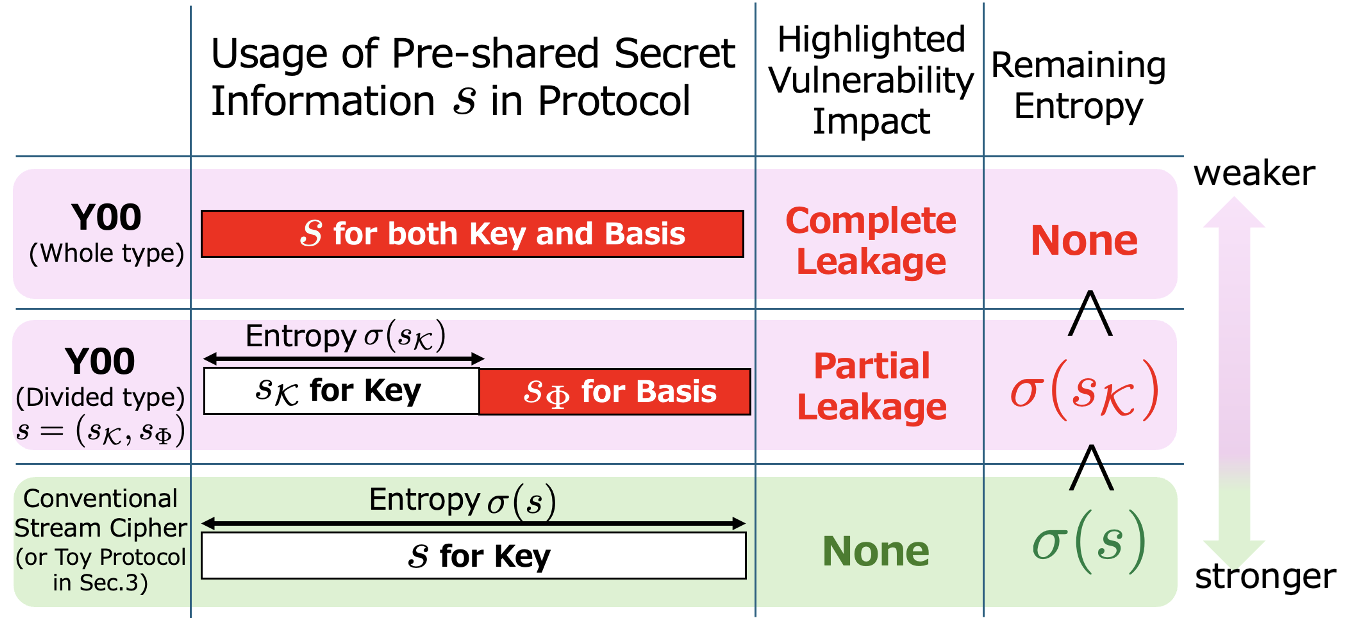}
\caption{Summary of main claim.\label{fig:main claim}}
\end{center}
\end{figure}

Fig.~\ref{fig:main claim} summarizes the impact of the structural vulnerability identified in the Y00 protocol. The Remaining Entropy is the entropy of the secret information left for legitimate users after the attack exploiting this vulnerability. If attackers supplement this attack with a known-plaintext attack equivalent to the Remaining Entropy, they can fully determine the pre-shared secret $s$. (An outline of known-plaintext attacks is provided in Sec.~\ref{sec:pre0}.) Since the amount of known plaintext required matches the Remaining Entropy, this figure illustrates that Y00 is susceptible to attack with less known plaintext than traditional stream ciphers. (The Whole type and Divided type in the figure represent variations of the Y00 protocol, explained in Sec.~\ref{sec:Y00 introduction}.) Thus, this paper concludes that Y00's structural vulnerability not only fails to enhance security but could actually undermine it.

To elucidate these findings, this paper is structured as follows:  
In Sec. \ref{sec:preliminaries}, we introduce conventional stream cipher and known-plaintext attacks as foundational concepts, followed by an overview of the Y00 protocol and its "dual structure" proposed for security. We then formalize their security claims, replacing vague and intuitive arguments with a clear logical framework to evaluate their achievements and limitations.
In Sec. \ref{sec:toy_protocol}, we introduce a toy protocol designed as a reference model. This protocol highlights the intrinsic vulnerabilities of Y00 and demonstrates that its claimed quantum properties do not contribute to meaningful security. The toy protocol also serves as a benchmark for comparing Y00's structural weaknesses against a simpler, more robust design.
In Sec. \ref{sec:vulnerability}, we analyze the structural vulnerability of Y00 in detail. This section begins with an intuitive explanation of the vulnerability, followed by a formal analysis using maximum likelihood estimation \cite{cover06}. By assuming PRNGs without intrinsic weaknesses, we isolate the fundamental design flaws of Y00 and demonstrate their impact on the protocol's security.
Finally, in Sec. \ref{sec:conclusion}, we conclude by discussing the broader implications of these vulnerabilities for practical information security. This includes recommendations for evaluating cryptographic protocols that incorporate physical-layer elements, ensuring they meet stringent security requirements.

Through this examination, this paper aims to shed light on significant overlooked security vulnerabilities in the Y00 protocol.

\section{Preliminaries}
\label{sec:preliminaries}
\subsection{Conventional Stream Cipher and Known-Plaintext Attacks}
\label{sec:pre0}
\begin{figure}[ht]
\centering
\begin{minipage}[b]{.5\linewidth}
\centering
\includegraphics[width=.9\linewidth]{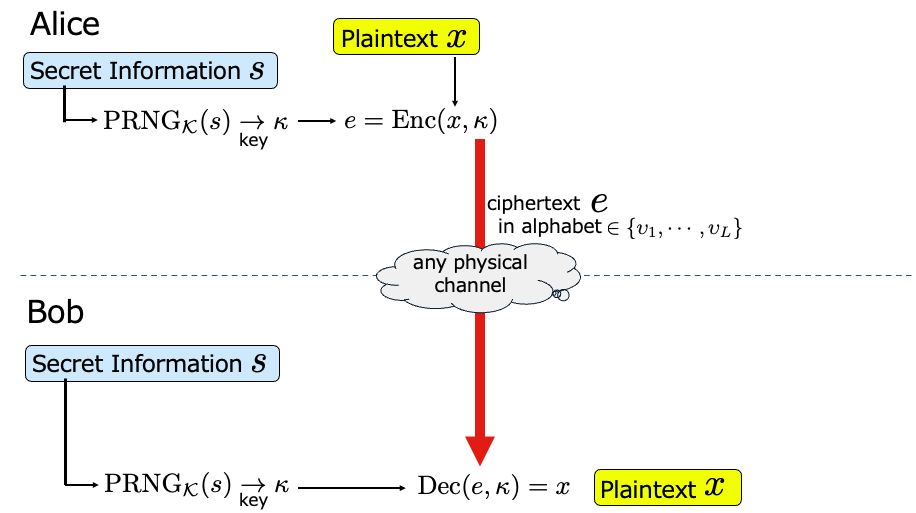}%
\subcaption{Conventional stream cipher}\label{fig:stream cipher protocol}%
\end{minipage}%
\hfil
\begin{minipage}[b]{.5\linewidth}
\centering
\includegraphics[width=.9\linewidth]{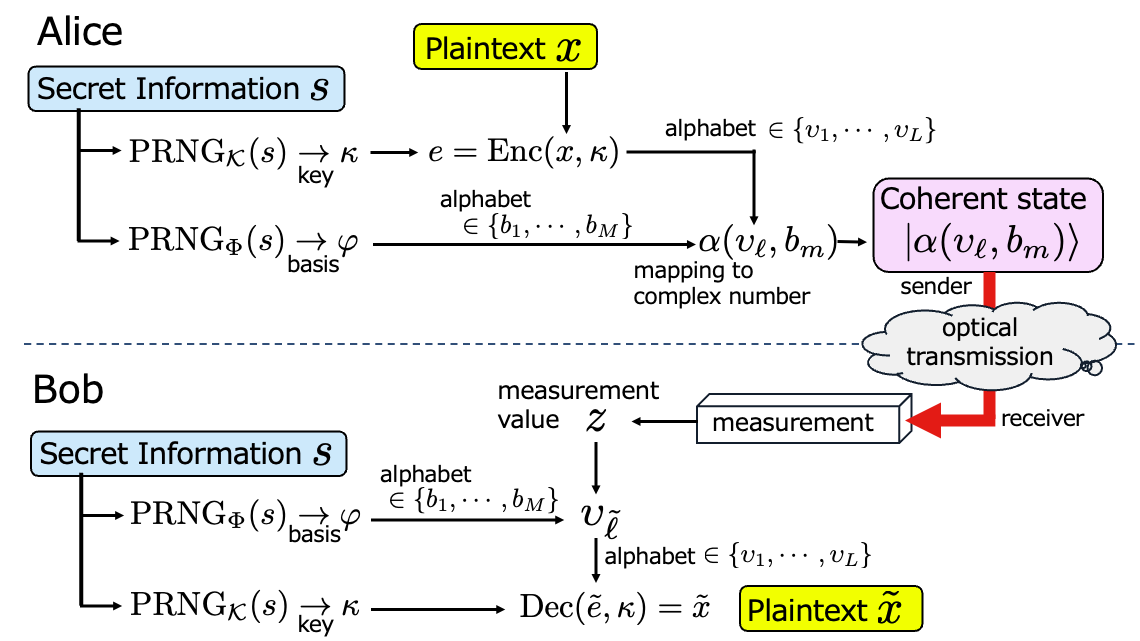}%
\subcaption{Y00 protocol}\label{fig:Y00 protocol}%
\end{minipage}%
\caption{Two protocols}
\label{fig_sim}
\end{figure}

This section introduces the basic operations of conventional stream ciphers, focusing on the aspects relevant to understanding its vulnerabilities in the context of this study.
Typically, as illustrated in Fig. \ref{fig:stream cipher protocol}, conventional stream cipher is conducted as follows:
\begin{enumerate}
    \item {\bf For both Alice and Bob}: 
    \begin{enumerate}
        \item Alice and Bob share secret information $s \in \mathcal{S}$ in advance. Here, $s$ is assumed to be uniformly selected from $\mathcal{S}$, giving the entropy of $s$ as
        \begin{equation}\label{eq:entropy of secret information}
        \sigma(s) = \log |{\mathcal S}|.
        \end{equation}
        \item Alice and Bob each generate a key $\kappa \in \mathcal{K}$ from $s$. This key is generated using a shared pseudorandom number generation (PRNG) algorithm:
        \[
        \mbox{PRNG}_{\mathcal K}:\mathcal{S}\mapsto\mathcal{K}
        \]
        such that
        \[
        \kappa=\mbox{PRNG}_{\mathcal K}(s).
        \]
    \end{enumerate}
    \item {\bf On Alice's side}:
    \begin{enumerate}
        \item Alice encrypts the plaintext $x \in \mathcal{X}$ using the key $\kappa$ to produce a ciphertext $e \in \mathcal{E}$, which she then transmits. The ciphertext is generated using an encryption algorithm,
        \[
        \mbox{Enc}:\mathcal{X}\times{\mathcal{K}}\mapsto\mathcal{E},
        \]
        such that
        \[
        e=\mbox{Enc}(x,\kappa).
        \]
    \end{enumerate}
    \item {\bf On Bob's side}:
    \begin{enumerate}
        \item Bob decrypts $x$ from the received $e$ and $\kappa$. This process uses a decryption algorithm
        \[
        \mbox{Dec}:\mathcal{E}\times{\mathcal{K}}\mapsto \mathcal{X},
        \]
        such that
        \begin{equation}\label{eq:decryption of classical streamcipher}
        x=\mbox{Dec}(e,\kappa).
        \end{equation}
    \end{enumerate}
\end{enumerate}

In a standard security analysis, we assume that an attacker (hereafter referred to as Eve) has access to all information except for the secret information. Thus, we assume that Eve, like Alice and Bob, has access to the pseudorandom number generator (PRNG) algorithm, the encryption algorithm $\mbox{Enc}$, and the decryption algorithm $\mbox{Dec}$ (or, conversely, define any information advantage of Alice and Bob over Eve as part of the secret).

In this context, if Eve learns the secret $s$ through some means, Alice and Bob lose their advantage over Eve and can no longer perform secure communication. Therefore, Eve's target in an attack may be not the message itself, but the pre-shared secret information $s$.

The known-plaintext attack (KPA) is one of the most well-known attacks in symmetric key cryptography, which includes conventional stream ciphers.
In this type of attack, an adversary exploits both a known plaintext and its corresponding ciphertext to deduce the secret information or encryption key. Here, Eve obtains both a plaintext $x'$ and the corresponding ciphertext $e'=\mbox{Enc}(x',\kappa)$ (for instance, by observing ciphertexts over the communication channel). Eve can then take a candidate secret $\hat{s} \in \mathcal{S}$, calculate a candidate key ${\kappa}(\hat{s}) := \mbox{PRNG}(\hat{s})$, and then calculate the candidate ciphertext
\[
{e'}(\hat{s}) := \mbox{Enc}(x',{\kappa}(\hat{s})),
\]
searching for a candidate $\hat{s}$ such that ${e'}(\hat{s})=e'$. Once $s$ is discovered, Eve can decrypt all ciphertexts as long as Alice and Bob continue to use the same shared secret $s$.

To provide further context for later discussions, let us analyze the mechanics of a known-plaintext attack in more detail. Given the binary representations of plaintext $x'$ and key $\kappa$:
\[
x'=(x'_1,\cdots,x'_N),\quad x'_n \in \{0,1\},
\]
\[
\kappa=(\kappa_1,\cdots,\kappa_N),\quad \kappa_n \in \{0,1\},
\]
we can adopt a specific form of $e'=\mbox{Enc}(x',\kappa)$:
\begin{equation}\label{eq:conventional ciphertext}
e'=(x'_1 \oplus \kappa_1,\cdots,x'_N \oplus \kappa_N),
\end{equation}
where $\oplus$ denotes the XOR operation. Under this assumption, the candidate ciphertext $e'(\hat{s})$ for a guessed secret $\hat{s}$ takes the form
\begin{equation}\label{eq:conventional ciphertext candidate}
e'(\hat{s})=(x'_1 \oplus \kappa_1(\hat{s}),\cdots,x'_N \oplus \kappa_N(\hat{s})),
\end{equation}
where
\[
\kappa(\hat{s})=(\kappa_1(\hat{s}),\cdots,\kappa_N(\hat{s}))
\]
is the binary representation of $\kappa(\hat{s})$. Eve's task in the known-plaintext attack is thus to identify $\hat{s}$ for which $\kappa_n = \kappa_n(\hat{s})$ for all $n$.

With a PRNG that behaves ideally for key generation, we have
\[
|{\mathcal S}_{\kappa_n}^{(n)}|\approx|{\mathcal S}_{\kappa_n \oplus 1}^{(n)}|\approx|{\mathcal S}|/2,
\]
where each set ${\mathcal S}_{\kappa_n}^{(n)}$ and ${\mathcal S}_{\kappa_n \oplus 1}^{(n)}$ is defined as
\[
{\mathcal S}_{\kappa_n}^{(n)} := \{\hat{s}~|~\hat{s}\in\mathcal{S}~\mbox{and}~\kappa_n(\hat{s})=\kappa_n\},
\]
\[
{\mathcal S}_{\kappa_n \oplus 1}^{(n)} := \{\hat{s}~|~\hat{s}\in\mathcal{S}~\mbox{and}~\kappa_n(\hat{s})=\kappa_n \oplus 1\}.
\]
The number of plaintext bits required to uniquely determine $s$ is approximately $\log_2 |{\mathcal S}|$, which directly relates to the entropy of $s$.

\subsection{Overview of the Y00 Protocol and Recent Claims}
\label{sec:Y00 introduction}

In the previous section, we reviewed conventional stream cipher and the associated known-plaintext attack (KPA). As discussed, the key aspect of KPA is that the attacker gains access to both the known plaintext and its corresponding ciphertext. This observation might lead to the idea that, if the ciphertext can be designed to be indistinguishable from other ciphertexts to the attacker, it might be possible to construct a symmetric key cryptosystem that is secure against KPA.
In fact, recent proposals related to the Y00 protocol have focused on realizing this idea by employing a "dual structure" approach, wherein the conventional ciphertext is generated and then transmitted as a quantum state. In this section, we provide an overview of the Y00 family of protocols and summarize their main claims, which are necessary for understanding the discussions in this paper.

The Y00 family of protocols, despite some variations in the use of quantum states, can generally be described as shown in Fig.~\ref{fig:Y00 protocol}, and can be outlined following the same steps as those described in the previous section, from Eq.~(\ref{eq:entropy of secret information}) to Eq.~(\ref{eq:decryption of classical streamcipher}):

\begin{enumerate}
    \item {\bf Common steps for Alice and Bob}:
    \begin{enumerate}
        \item Alice and Bob share a secret $s \in \mathcal{S}$ in advance.
        \item From the shared secret $s$, Alice and Bob independently generate a key $\kappa \in \mathcal{K}$ using a shared pseudorandom number generator (PRNG):
        \begin{equation}\label{eq:PRNG00}
        \mbox{PRNG}_{\mathcal K}:\mathcal{S}\mapsto\mathcal{K},
        \end{equation}
        \begin{equation}\label{eq:PRNG01}
        \kappa=\mbox{PRNG}_{\mathcal K}(s).
        \end{equation}
        \item From the shared secret $s$, Alice and Bob also generate a sequence of bases $\varphi \in \Phi$ using a separate PRNG algorithm:
        \begin{equation}\label{eq:PRNG10}
        \mbox{PRNG}_\Phi:\mathcal{S}\mapsto\Phi,
        \end{equation}
        \begin{equation}\label{eq:PRNG11}
        \varphi=\mbox{PRNG}_\Phi(s).
        \end{equation}
        The sequence $\varphi$ can be represented as:
        \begin{equation}\label{eq:key for y00}
        \varphi = \{\varphi_1, \cdots, \varphi_n, \cdots, \varphi_N\},
        \end{equation}
        where each $\varphi_n$ is an element from the set of bases:
        \begin{equation}\label{eq:alphabet of basis}
        \mathcal{B} := \{b_1, \cdots, b_m, \cdots, b_M\}.
        \end{equation}
    \end{enumerate}
    \item {\bf On Alice's side}:
    \begin{enumerate}
        \item Alice generates a ciphertext $e \in \mathcal{E}$ from the plaintext $x \in \mathcal{X}$ and the key $\kappa$ using the encryption algorithm $\mbox{Enc}$:
        \[
        e=\mbox{Enc}(x, \kappa).
        \]
        The ciphertext $e$ is then transformed into a sequence of $L$ alphabet symbols:
        \begin{equation}\label{eq:representation of ciphertext}
        e \rightarrow \{e_1^{(L)}, \cdots, e_n^{(L)}, \cdots, e_N^{(L)}\},
        \end{equation}
        where each $e_n^{(L)}$ is an element from the set:
        \begin{equation}\label{eq:alphabet of ciphertext}
        \Upsilon := \{\upsilon_1, \cdots, \upsilon_\ell, \cdots, \upsilon_L\}.
        \end{equation}
        \item For each $n = 1, \ldots, N$, Alice transmits to Bob the quantum state determined by $e_n^{(L)} = \upsilon_\ell$ and the basis information $\varphi_n = b_m$:
        \begin{equation}\label{eq:general Y00 state}
        \rho_{\upsilon_\ell, b_m}.
        \end{equation}
    \end{enumerate}
    \item {\bf On Bob's side}:
    \begin{enumerate}
        \item For each $n$, Bob performs a quantum measurement on the received quantum state $\rho_{\upsilon_\ell, b_m}$, obtaining the measurement result $z_n$. Using the shared basis information $\varphi_n = b_m$, Bob estimates the value of $e_n^{(L)}$ as $\tilde{e}_n^{(L)}$.
        \item The sequence $\{\tilde{e}_1^{(L)}, \cdots, \tilde{e}_N^{(L)}\}$ is then reassembled into $\tilde{e}$:
        \begin{equation}\label{eq:representation of ciphertext2}
        \{\tilde{e}_1^{(L)}, \cdots, \tilde{e}_N^{(L)}\} \rightarrow \tilde{e},
        \end{equation}
        which is decrypted using the decryption algorithm $\mbox{Dec}$:
        \begin{equation}\label{eq:decryption}
        \tilde{x} = \mbox{Dec}(\tilde{e}, \kappa),
        \end{equation}
        yielding an estimate of the plaintext $\tilde{x}$.
    \end{enumerate}
\end{enumerate}

It is important to note that the "dual structure" of the Y00 protocol involves two separate PRNG sequences shared between Alice and Bob: the first sequence, $\kappa$, is used for standard encryption and decryption, while the second sequence, $\varphi$, determines the quantum state used as the carrier. If the protocol directly encodes $x$ into the quantum state (as originally proposed), steps 1-(b), 2-(a), and 3-(b) could be omitted with $e = x$. 

It should be noted that there is some ambiguity in the choice of the domain for the pseudorandom number generators $\mbox{PRNG}_{\mathcal K}$ and $\mbox{PRNG}_{\Phi}$ used in steps 1-(b) and 1-(c). One approach, as described in Eqs.~(\ref{eq:PRNG00})--(\ref{eq:PRNG11}), is to use the same domain $\mathcal S$ for both generators and to derive different random sequences from the same secret information $s$. We shall refer to this as the \textit{whole type} approach. Another method is to partition the domain as ${\mathcal S}={\mathcal S}_{\mathcal K}\times{\mathcal S}_{\Phi}$, and split the secret $s$ into $s = (s_{\mathcal K}, s_{\Phi}) \in {\mathcal S}_{\mathcal K}\times{\mathcal S}_{\Phi}$, allowing for separate PRNGs as follows:
\begin{equation}\label{eq:devided0}
\mbox{PRNG}_{\mathcal K}: {\mathcal S}_{\mathcal K}\mapsto {\mathcal K}, \quad \kappa=\mbox{PRNG}_{\mathcal K}(s_{\mathcal K}),
\end{equation}
\begin{equation}\label{eq:devided1}
\mbox{PRNG}_{\Phi}: {\mathcal S}_{\Phi}\mapsto \Phi, \quad \varphi=\mbox{PRNG}_{\Phi}(s_{\Phi}).
\end{equation}
We will refer to this as the \textit{divided type} approach. It should be noted that this distinction is not explicitly addressed in the proponents' papers, and there are no clear descriptions regarding the choice of approach. However, the vulnerabilities identified in this paper exhibit different impacts depending on the type used. We will revisit this issue in later sections.

For Alice and Bob to successfully share the message $x$, the condition that must be met is that the estimated value $\tilde{e}$, derived by Bob from the quantum measurement results and the basis information $\varphi$, should be equal to the ciphertext $e$ generated by Alice, with a negligible error probability. The quantum states defined in Eq.~(\ref{eq:general Y00 state}) are designed such that this condition holds for Bob, who shares the basis information $\varphi$ with Alice. Conversely, these quantum states aim to prevent Eve, who does not have access to the basis information, from making a correct estimation. According to recent claims in the literature, this design is believed to prevent Eve from obtaining the ciphertext $e$ itself, even when she has partial knowledge of the plaintext, supposedly making known-plaintext attacks challenging or even infeasible.

\subsection{Choice of the Quantum State}
\label{sec: choice of state}
In this section, we describe the specific form of the quantum state in Eq. (\ref{eq:general Y00 state}) as relevant to the objectives of this study.

The Y00 family of protocols consistently employs optical coherent states \cite{scully97} as quantum carriers. Optical coherent states are relatively stable and well-suited for communication systems. A coherent state corresponds one-to-one with a complex number $\alpha \in \mathbb{C}$ and is represented by the state $|\alpha\rangle$. The fidelity between two coherent states $|\alpha\rangle$ and $|\alpha'\rangle$ is given by:
\[
|\langle \alpha'|\alpha\rangle|^2 = \exp\left(-|\alpha-\alpha'|^2\right).
\]
This indicates that when $|\alpha - \alpha'| \gg 1$, the two states are nearly orthogonal and thus highly distinguishable. Conversely, when $|\alpha - \alpha'| \ll 1$, the overlap is substantial, making the states difficult to distinguish. In particular, when a heterodyne measurement, which is a type of POVM measurement defined by:
\begin{equation}\label{eq:heterodyne measurement}
\left\{\frac{1}{\pi} |z\rangle\langle z|\right\}_{z\in\mathbb{C}},
\end{equation}
is applied to a coherent state $|\alpha\rangle$, the resulting measurement outcome $z$ follows a probability density distribution given by:
\begin{equation}\label{eq:heterodyne measurement prob}
p_\alpha(z) = \frac{1}{\pi} \exp\left(-|z-\alpha|^2\right).
\end{equation}
Thus, distinguishing between the two states $|\alpha\rangle$ and $|\alpha'\rangle$ based on this measurement reduces to a decision problem of whether $z$ was sampled from $p_\alpha(z)$ or $p_{\alpha'}(z)$. In essence, when $|\alpha - \alpha'| \gg 1$, the two states can be easily distinguished using heterodyne measurement. In contrast, when $|\alpha - \alpha'| \ll 1$, distinguishing them becomes significantly challenging.

Leveraging this distinguishability property of coherent states, the Y00 family of protocols implements the quantum states in Eq. (\ref{eq:general Y00 state}) using coherent states $|\alpha(\upsilon_\ell, b_m)\rangle$. Specifically, a mapping function:
\[
\alpha: \Upsilon\times\mathcal{B} \mapsto \mathbb{C},
\]
is defined to specify the coherent state used in the protocol.

Several mapping strategies based on different mapping functions have been proposed in the literature. However, all of these mappings satisfy the condition for Bob, who shares the basis information, to correctly estimate the ciphertext with minimal error:
\begin{equation}\label{eq:mapping rule for correctness condition}
\forall b_m \in \mathcal{B}, |\alpha(\upsilon_\ell,b_m)-\alpha(\upsilon_{\ell'},b_m)|\gg1, \quad\text{for }\upsilon_\ell\neq\upsilon_{\ell'}.
\end{equation}

By comparison, the requirement for Eve, who lacks the basis information, differs, to be unable to correctly estimate the ciphertext varies among different proposals. Two main variations exist:
\begin{itemize}
    \item One condition, achieved even in the original Y00 proposal, is as follows:
    \begin{equation}\label{eq:hiding condition00}
    \left[
    \begin{array}{c}
    \forall (\upsilon_\ell, b_m) \in \Upsilon\times\mathcal{B}, ~\forall \upsilon_{\ell'}\in \Upsilon, ~\exists b_{m'}\in\mathcal{B} \\
    \text{s.t.}\quad |\alpha(\upsilon_\ell, b_m) - \alpha(\upsilon_{\ell'}, b_{m'})| \ll 1,
    \end{array}\right]
    \end{equation}
    and
    \begin{equation}\label{eq:hiding condition01}
    \left[
    \begin{array}{c}
    \text{The number of } b_{m'} \text{ satisfying Eq. (\ref{eq:hiding condition00}) is}\\
    \text{approximately independent of }\\
     \upsilon_\ell, b_m \text{, and } \upsilon_{\ell'}.
    \end{array}\right]
    \end{equation}
    The condition in Eq. (\ref{eq:hiding condition00}) aims to prevent Eve, who lacks the basis information $b_m$, from excluding any possible value of $\upsilon_{\ell'}$. Moreover, Eq. (\ref{eq:hiding condition01}) ensures uniformity across different $\upsilon_{\ell'}$ values. This condition likely stems from the initial design of Y00, where $x$ was directly encoded into the quantum state without encryption, and thus, minimizing any leakage of $x$ was a priority. Examples of proposals that adopt this condition include those utilizing only the phase direction of the complex plane \cite{barbosa03,banwell05,tanizawa19}, those relying solely on amplitude modulation \cite{hirota05}, and those leveraging a lattice structure \cite{kato05}. For subsequent discussions, we refer to the most representative phase-based approach as the P-type, illustrated in Fig. \ref{fig:y00 phase type}.
\begin{figure}[ht]
\begin{minipage}[b]{0.5\linewidth}
\centering
\includegraphics[width=.9\linewidth]{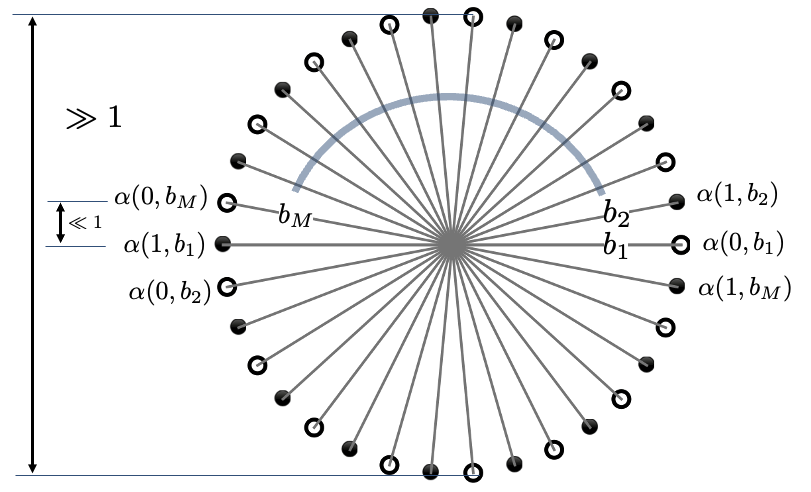}%
\subcaption{P-type mapping with $(M,L)=(17,2)$}\label{fig:y00 phase type}%
\end{minipage}%
\hfil
\begin{minipage}[b]{.5\linewidth}
\centering
\includegraphics[width=.9\linewidth]{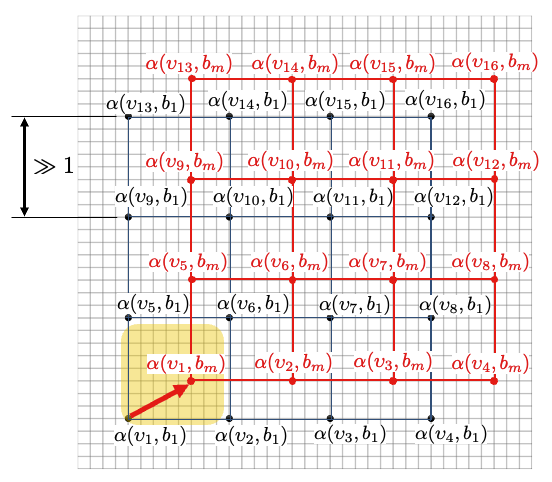}%
\subcaption{N-type mapping with $(M,L)=(64,16)$}\label{fig:y00 nakazawa type}%
\end{minipage}%
\caption{Example of mapping functions}
\label{fig_mapping}
\end{figure}

\item Another protocol, which satisfies a different set of conditions, was proposed by Nakazawa et al. \cite{nakazawa14}. The condition satisfied by this protocol is given by:
\begin{equation} \label{eq:hiding condition10}
\left[
\begin{array}{c}
\forall (\upsilon_\ell, b_m) \in \Upsilon \times \mathcal{B}, ~\exists (\upsilon_{\ell'}, b_{m'}) \in \Upsilon \times \mathcal{B}\\
\mbox{s.t.}~ |\alpha(\upsilon_\ell, b_m) - \alpha(\upsilon_{\ell'}, b_{m'})| \ll 1~~\mbox{and}~~ \\ \upsilon_{\ell'}\not=\upsilon_\ell
\end{array}\right]
\end{equation}

It should be noted that the condition in Eq. (\ref{eq:hiding condition10}) appears to be significantly relaxed compared to the condition in Eq. (\ref{eq:hiding condition01}), as it requires only the existence of at least one $\upsilon_{\ell'}\neq\upsilon_\ell$ that cannot be ruled out. In fact, if the condition in Eq. (\ref{eq:hiding condition01}) is satisfied, then the condition in Eq. (\ref{eq:hiding condition10}) is also satisfied; however, the converse does not hold. In other words, this condition allows for a certain degree of leakage of $\upsilon_\ell$ information based on the measurement result $z$. For proponents of the effectiveness of the previously mentioned "dual structure," the difference between the conditions in Eq.  (\ref{eq:hiding condition01}) and (\ref{eq:hiding condition10}) should carry significant implications. However, neither the original paper \cite{nakazawa14} nor subsequent experimental reports \cite{yoshida15, yoshida16, tan20, chen21, zhang22, hanwen24} have explicitly addressed this difference, and it remains unclear whether the authors were aware of this distinction.

An example of the mapping proposed in \cite{nakazawa14} is illustrated in Fig. \ref{fig:y00 nakazawa type}. (In this paper, we refer to this scheme as the N-type for convenience.) First, by fixing $b_1$, different $\upsilon_\ell$ values are arranged as lattice points on the complex plane, forming a configuration denoted by $\mathcal{G}(b_1)$, which is defined as follows:
$$
\mathcal{G}(b_1) := \{\alpha(\upsilon_\ell, b_1)\}_{\ell \in \{1, \cdots, L\}}.
$$
In the figure, $\mathcal{G}(b_1)$ corresponds to the black lattice. Note that, according to the condition in Eq. (\ref{eq:mapping rule for correctness condition}), each lattice point in this configuration is sufficiently separated. Similarly, we introduce $\mathcal{G}(b_{m})$, represented by the red lattice in the figure, which is a parallel translation of $\mathcal{G}(b_1)$ on the complex plane. The direction and magnitude of this translation depend on the value of $b_m$, and the design involves selecting one of the 64 small lattice points in the yellow region at the lower left corner of the figure. The maximum displacement is approximately the size of a single lattice unit, and for neighboring lattice points $\upsilon_\ell$ and $\upsilon_{\ell'}$, it becomes difficult to distinguish them from the measurement results unless the value of $b_m$ is known. This design ensures that the condition in Eq. (\ref{eq:hiding condition10}) is satisfied.
\end{itemize}

From the preceding discussion, the Y00 protocol may initially appear to be an appealing proposal that integrates quantum states into traditional cryptographic techniques to enhance physical security. However, in reality, the quantum mechanical effects do not play a central role; rather, they merely serve as a superficial feature that gives an impression of leveraging advanced technology. Worse still, the Y00 protocol possesses inherent vulnerabilities stemming from its structural design. These issues will be elaborated upon in the subsequent sections.

\section{Toy Protocol}
\label{sec:toy_protocol}

In this section, we introduce a Toy protocol that serves as a reference for understanding both the illusory quantum effects and the inherent vulnerabilities of the Y00 protocol. The Toy protocol aims to provide a comparative framework to elucidate the structural issues of Y00.

The Toy protocol is defined in a manner almost identical to the Y00 protocol described in equations (\ref{eq:PRNG00}) through (\ref{eq:decryption}), with the following differences:

\begin{enumerate}
    \item For simplicity, we set the alphabet set $\mathcal{B}$ in Eq. (\ref{eq:alphabet of basis}) to $M = 2$, with $b_1 = 0$ and $b_2 = 1$. Similarly, the alphabet set $\Upsilon$ in Eq. (\ref{eq:alphabet of ciphertext}) is set to $L = 2$, with $\upsilon_1 = 0$ and $\upsilon_2 = 1$.
    \item We assume a specific form for the state $\rho_{\upsilon_\ell, b_m}$ in Eq. (\ref{eq:general Y00 state}):
    \begin{equation}\label{eq:states for toy}
        \rho_{\upsilon_\ell, b_m} = \nu_{\upsilon_\ell \oplus b_m},
    \end{equation}
    where, due to the restrictions on $\mathcal{B}$ and $\Upsilon$, the states $\nu_{\upsilon_\ell \oplus b_m}$ are limited to just two distinct states, $\nu_0$ and $\nu_1$. These two states can be either quantum or classical, but we assume they are orthogonal and thus always distinguishable. In other words, even if the states are implemented using quantum systems, they are effectively equivalent to a classical system with no emergent quantum properties.
\end{enumerate}

It is straightforward to verify that the protocol defined above satisfies both the conditions in Eqs. (\ref{eq:mapping rule for correctness condition}) and  (\ref{eq:hiding condition01}). (Specifically, the condition "$|\alpha(\upsilon_\ell, b_m) - \alpha(\upsilon_{\ell'}, b_m)| \gg 1 \quad \text{for} \quad \upsilon_\ell \neq \upsilon_{\ell'}$" can be interpreted as "the states $\rho_{\upsilon_\ell, b_m}$ and $\rho_{\upsilon_{\ell'}, b_m}$ are distinguishable," and the condition "$|\alpha(\upsilon_\ell, b_m) - \alpha(\upsilon_{\ell'}, b_{m'})| \ll 1$" can be interpreted as "the states $\rho_{\upsilon_\ell, b_m}$ and $\rho_{\upsilon_{\ell'}, b_{m'}}$ are (identical and hence) indistinguishable.") Recall that if Eq. (\ref{eq:hiding condition01}) is satisfied, then Eq. (\ref{eq:hiding condition10}) is also satisfied. Thus, this Toy protocol satisfies the conditions claimed as the basis for the security of Y00. This implies that the claimed security does not rely on quantum effects.

Next, let us examine the essence of what they refer to as the "dual structure" by closely analyzing the meaning of the Toy protocol introduced above. One possible interpretation of this Toy protocol is that, depending on the value of $b_m$, the association between the two physical states ${\nu_0, \nu_1}$ and the two ciphertext values ${\upsilon_0, \upsilon_1}$ toggles as follows:
\begin{itemize}
\item
When $b_m=0$, the state $\nu_0$ corresponds to $\upsilon_0$ and $\nu_1$ corresponds to $\upsilon_1$.
\item
When $b_m=1$, the state $\nu_0$ corresponds to $\upsilon_1$ and $\nu_1$ corresponds to $\upsilon_0$.
\end{itemize}
This interpretation aligns with the idea of "additional protection of the ciphertext via physical states," which is claimed to be achieved by Y00's dual structure.

However, there is an alternative interpretation. One could take the XOR of the ciphertext value $\upsilon_\ell$ and $b_m$ (where $b_m$ is the output of the pseudo-random number generator $\varphi = \text{PRNG}_\Phi(s)$, as recalled earlier), effectively creating a doubly-encrypted value:
\[
\omega_{\ell m} := \upsilon_\ell \oplus b_m.
\]
This new ciphertext value $\omega_{\ell m}$ is then mapped consistently to the physical state $\nu_{\omega_{\ell m}}$. These are merely different interpretations of the same protocol; therefore, the proponents' claim of "additional protection of the ciphertext via physical states" amounts to nothing more than "double encryption." The advantage against attackers in such a double encryption scheme relies entirely on the entropy of the pre-shared secret information, indicating that the effect of physical states for additional protection is solely dependent on the pre-shared secret. (In scenarios where no specific vulnerabilities in the pseudorandom number generator are assumed, as in this paper, the security level of a conventional stream cipher using secret information $s$ is equivalent to that of the Toy protocol \cite{trappe05}. Thus, in practice, the double encryption scheme of the Toy protocol does not provide any additional security enhancement; using a conventional stream cipher instead of the Toy protocol would achieve the same level of security.)

It is important to reiterate that even if quantum states are employed as physical states, no new protection mechanism arising from quantum properties is introduced.
In this section, we have shown that the security claimed by the proponents of Y00 can be achieved by the Toy protocol, and that the actual mechanism of this security is double encryption. However, whether Y00 can achieve the same level of security as the Toy protocol is another matter, and it might be true that the Toy protocol offers stronger cryptographic protection. In fact, Y00 exhibits vulnerabilities that are absent in the Toy protocol, as we will discuss in the next section.

\section{Structural Vulnerability}
\label{sec:vulnerability}
In this section, we identify a vulnerability that arises from the inherent structure of the Y00 protocol. We begin by presenting an intuitive understanding of this vulnerability based on approximate reasoning. Following this, we refine the argument using maximum likelihood estimation\cite{cover06} to provide a more rigorous analysis. Additionally, we demonstrate that this vulnerability is absent in the Toy protocol introduced in the previous section, clarifying that the root cause of the issue lies specifically in the structural features unique to Y00.

For the purpose of clarity, we temporarily focus on the P-type Y00 protocol as illustrated in Fig.~\ref{fig:y00 phase type}. Nevertheless, we emphasize at the outset that the analysis provided here is equally applicable to all variations of Y00 proposed thus far, as will become evident in the subsequent discussion.

\subsection{Intuitive Explanation}
\label{eq:intuitive explanation}

We begin by intuitively understanding that the measurement outcome $z$ obtained by Eve may contain crucial information about the secret key $s$.

\begin{figure}[ht]
\begin{minipage}[b]{0.5\linewidth}
\centering
\includegraphics[width=0.9\linewidth]{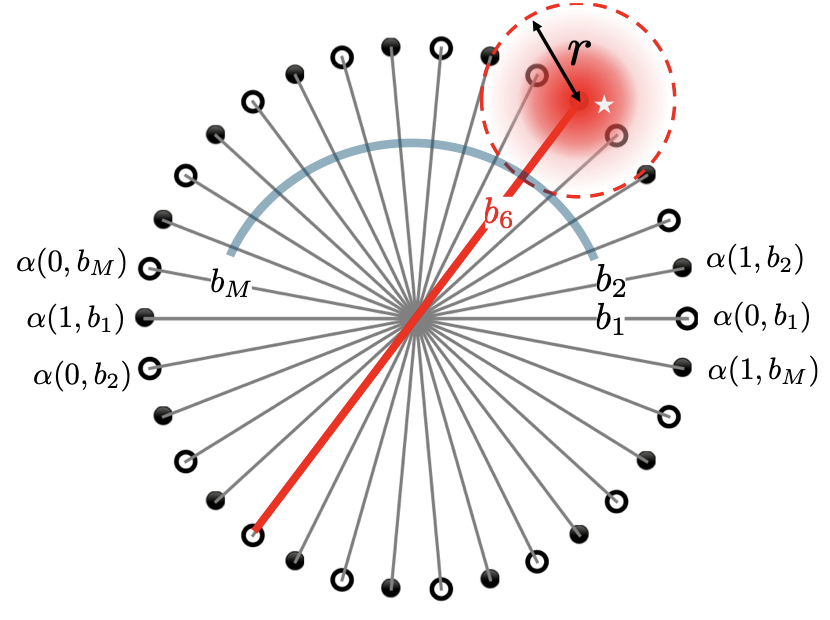}%
\subcaption{Measurement value}\label{fig:y00 view a}%
\end{minipage}%
\hfil
\begin{minipage}[b]{0.5\linewidth}
\centering
\includegraphics[width=0.9\linewidth]{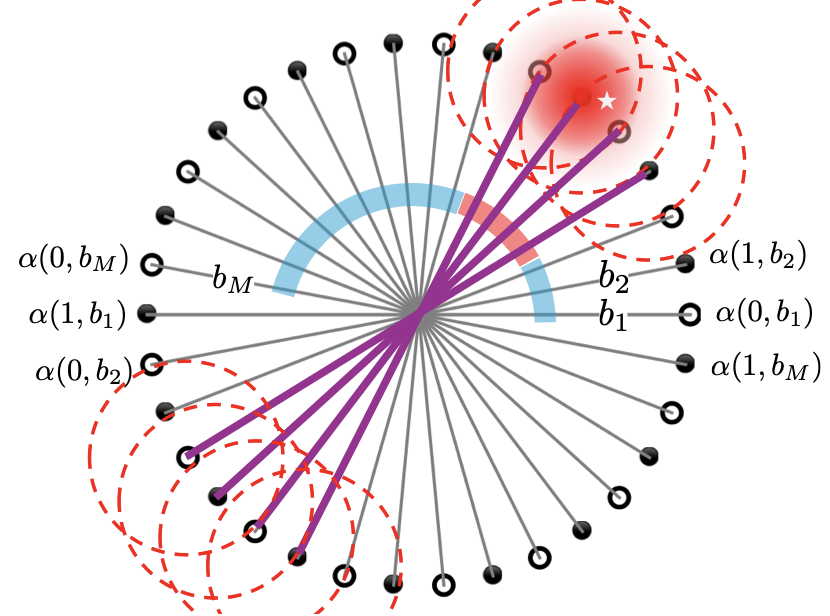}%
\subcaption{Example of ${\mathcal B}^{\pm}(z)$}\label{fig:y00 view b}%
\end{minipage}%
\caption{View from Eve}
\label{view from eve}
\end{figure}

Consider, for example, the scenario depicted in Fig.~\ref{fig:y00 view a}, where the basis $b_6$ is selected from the set of bases $\{b_1, \cdots, b_{16}\}$, and the quantum state $|\alpha(1, b_6)\rangle$ is used to transmit $\upsilon_2 = 1$. When Eve performs a heterodyne measurement as described by Eq. (\ref{eq:heterodyne measurement}), she obtains a measurement value $z \in \mathbb{C}$ that statistically follows the probability density function given by Eq. (\ref{eq:heterodyne measurement prob}):
\[
p_{\alpha(1, b_6)}(z) = \frac{1}{\pi} \exp\left(-|z-\alpha(1, b_6)|^2\right).
\]
In Fig.~\ref{fig:y00 view a}, this probability distribution is represented by the red shaded area, and the measurement value $z$ is indicated by the hollow star. Since Eve does not know which basis was used, she cannot distinguish whether $\upsilon_1 = 0$ or $\upsilon_1 = 1$ was transmitted, as described in the arguments presented in Sec.~\ref{sec:Y00 introduction}.

However, let us now shift our perspective and consider the possibility of estimating the basis from the measurement value $z$. While it is impossible to determine the basis uniquely from a single measurement, the Gaussian shape and localized nature of the probability distribution imply that certain bases are more consistent with the observed value $z$. As illustrated in Fig.~\ref{fig:y00 view b}, we can introduce two sets: ${\mathcal B}^{+}(z)$, the set of bases that are consistent with the measurement (shown in purple), and ${\mathcal B}^{-}(z):={\mathcal B}\setminus{\mathcal B}^{+}(z)$, the set of all other bases.

To utilize the localization of the probability distribution for an attack, consider a circle of radius $r$ that fully encloses the Gaussian distribution with variance $1$ on the complex plane. Furthermore, assume that the two points $\alpha(\upsilon_{\ell}, b_m)$ and $\alpha(\upsilon_{\ell'}, b_m)$, corresponding to different values of $\upsilon_{\ell}$, are separated by at least $2r$. Specifically, we assume that (approximately)
\begin{equation}\label{eq:approximation}
\int_0^{2\pi} d\theta \int_0^r \frac{1}{\pi} \exp(-r'^2) r' dr \simeq 1,
\end{equation}
and that
\begin{equation}
|\alpha(\upsilon_{\ell}, b_m) - \alpha(\upsilon_{\ell'}, b_m)| > 2r.
\end{equation}
In practice, if we consider the situation in Fig.~\ref{fig:y00 view a}, where $|\alpha| \ge 10$ (i.e., coherent states with an average photon number of at least $100$), setting $r = 10$ makes Eq. (\ref{eq:approximation}) hold with an error of the order of $10^{-44}$. If this error can be treated as negligible, then the following explanation can be considered not only intuitive but also accurate.

Using this $r$, we define
\begin{equation}\label{eq:B^plus}
{\mathcal B}^{+}(z) = \{b_m \mid \exists \upsilon_{\ell} \in \Upsilon, |z - \alpha(\upsilon_{\ell}, b_m)| < r \}, \quad b_m \in {\mathcal B}.
\end{equation}
In the example shown in Fig.~\ref{fig:y00 view b}, we have:
\[
{\mathcal B}^+(z) \!\!=\!\!\{b_4, b_5, b_6, b_7\}, ~{\mathcal B}^-(z)\!\!=\!\!\{b_1, b_2, b_3, b_8, \cdots, b_M\}.
\]
Although the members of ${\mathcal B}^{\pm}(z)$ may vary depending on the measurement outcome for the same quantum state $|\alpha(1, b_6)\rangle$, within the approximation of Eq. (\ref{eq:approximation}), it is guaranteed that $b_6 \in {\mathcal B}^+(z)$. If Eve repeats this process for each coherent state transmitted over the communication channel, denoting the $n$th measurement result as $z_n$, she obtains a corresponding set ${\mathcal B}^+(z_n)$ for each measurement. Recalling that the true basis sequence is given by Eq. (\ref{eq:key for y00}), we can conclude that (within the approximation range of Eq. (\ref{eq:approximation})) $\varphi_n \in {\mathcal B}^+(z_n)$ holds for all $n$.

Using the reasoning outlined above, it is possible for Eve to determine the secret information $s$. Recall that $\varphi = \mbox{PRNG}_\Phi(s)$ as defined by Eq. (\ref{eq:PRNG11}). Eve can generate a candidate sequence of bases $\varphi(\hat{s}) := \mbox{PRNG}_\Phi(\hat{s}) = (\varphi_1(\hat{s}), \cdots, \varphi_n(\hat{s}), \cdots, \varphi_N(\hat{s}))$ for each candidate $\hat{s} \in \mathcal{S}$. The true secret $s$ can then be identified as the candidate $\hat{s}$ for which $\varphi_n(\hat{s}) \in {\mathcal B}^+(z_n)$ holds for all $n$.

If $\hat{s} \neq s$, then $\varphi_n(\hat{s})$ should appear uniformly distributed over $\mathcal B$ and uncorrelated with $\varphi_n(s)$. Therefore, the proportion of candidates $\hat{s}$ for which $\varphi_n(\hat{s}) \in {\mathcal B}^+(z_n)$ holds is approximately:
\begin{equation}\label{eq:consistent basis ratio}
R(z_n) := \frac{|{\mathcal B}^+(z_n)|}{M} \le 1.
\end{equation}

The proportion of candidates $\hat{s}$ that satisfy $\varphi_n(\hat{s}) \in {\mathcal B}^+(z_n)$ for all $N$ measurements is then given by:
\begin{equation}
\prod_{n=1}^{N} R(z_{n}).
\end{equation}

This quantity decreases exponentially with $N$ unless:
\begin{equation}
\forall z_n, \quad R(z_n) = 1.
\end{equation}

As long as there exists a region where $R(z) < 1$, the number of candidates decreases as Eve collects more measurements, eventually allowing her to uniquely determine the true secret $s$.

\begin{figure}[ht]
\begin{minipage}[b]{.5\linewidth}
\centering
\includegraphics[width=0.9\linewidth]{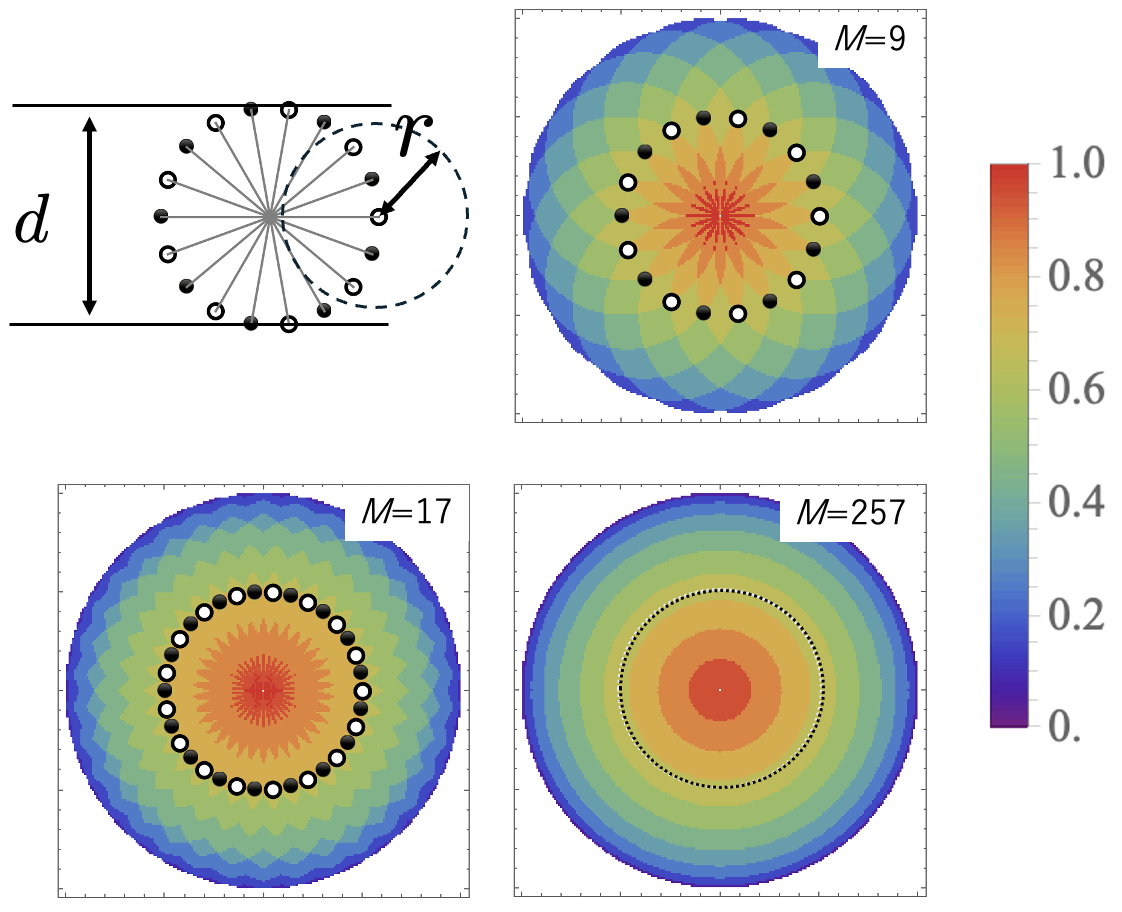}%
\subcaption{P-type with $r/d=1/2$}\label{fig:rz phase type}%
\end{minipage}%
\hfill
\begin{minipage}[b]{.5\linewidth}
\centering
\includegraphics[width=0.9\linewidth]{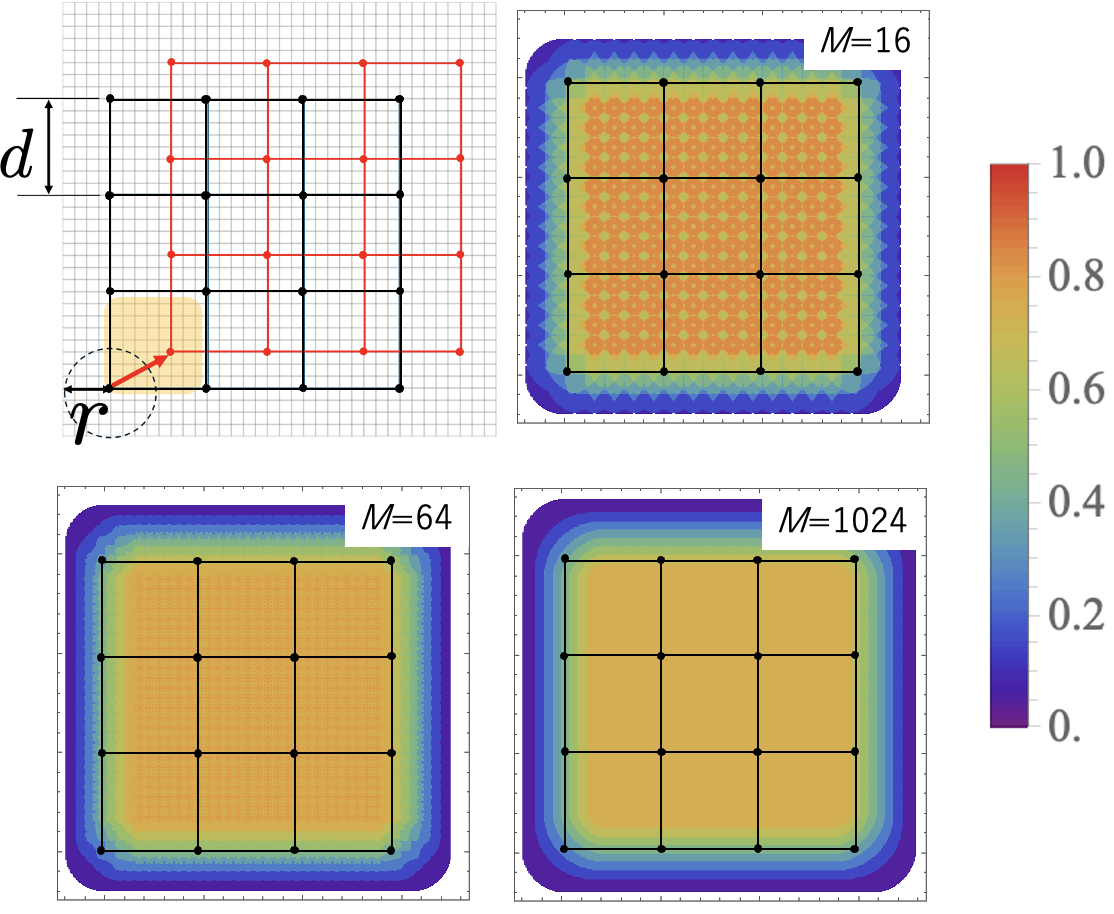}%
\subcaption{N-type with $r/d=1/2$}\label{fig:rz nakazawa type}%
\end{minipage}%
\caption{Numerical Plot of $R(z)$}
\label{fig:rz plot}
\end{figure}

Let us examine the numerical behavior of $R(z)$. Fig. \ref{fig:rz phase type} shows the plot of $R(z)$ for the P-type Y00 protocol (as in Fig. \ref{fig:y00 phase type}), and Fig. \ref{fig:rz nakazawa type} shows the corresponding plot for the N-type Y00 protocol (as in Fig. \ref{fig:y00 nakazawa type}). Both figures assume the worst-case scenario for the attacker, with a fixed ratio of $r/d = 1/2$. Note that, given a specific value of $r/d$, the plots can be generated independently of the value of $|\alpha|$.

In both cases, the region where $R(z) \lneq 1$ dominates, indicating that the attack described above is effective. An additional noteworthy observation is that increasing $M$ beyond a certain threshold does not significantly change the $z$-dependence of $R(z)$. This can be understood from the fact that as $M$ increases, $|{\mathcal B}^+(z)|$ grows proportionally with $M$. Therefore, merely increasing $M$ does not enhance resistance against this attack. Existing studies that have positively evaluated the security of Y00 have claimed that increasing $M$ enhances security by making adjacent states harder to distinguish. Although this may seem contradictory, it is merely due to their oversight regarding the specific leakage mechanism discussed in this paper, rather than any deeper reason.

The narrowing mechanism in this attack may appear similar to the narrowing process in a known-plaintext attack, as introduced in Sec.~\ref{sec:pre0}. However, it is important to emphasize that the current attack requires no known plaintext. In conventional stream ciphers, estimating the secret information $s$ necessitates a known plaintext whose entropy matches the entropy $\sigma(s)$ of the pre-shared secret. In contrast, the above attack achieves narrowing without requiring any known plaintext, revealing a structural vulnerability in Y00 that is absent in typical stream ciphers.

The discussion thus far pertains to the case where the basis sequence $\varphi$ depends entirely on the secret information $s$, as expressed in Eq.~(\ref{eq:key for y00}), which corresponds to the \textit{whole type} introduced in Sec.~\ref{sec:Y00 introduction}. For the \textit{divided type}, as introduced in Eqs.~(\ref{eq:devided0}) and~(\ref{eq:devided1}), the above attack only exposes the part of the secret information contributing to the generation of $\varphi$, specifically $s_\Phi$. In this sense, the \textit{divided type} can serve as a partition, preventing the complete leakage of $s$ under this attack. However, the amount of known plaintext required to reveal the remaining secret $s_{\mathcal K}$ corresponds to the entropy of $s_{\mathcal K}$, which means there is still a reduction by the entropy of $s_\Phi$ compared to a conventional stream cipher that uses the entire $s$ for key generation from the outset. As summarized in Fig.~\ref{fig:main claim}, it is worth emphasizing that whether using the \textit{whole type} or the \textit{divided type}, the implementation of a dual structure in Y00 not only fails to enhance security but rather introduces a new vulnerability.

The essence of the attack in this section lies in the fact that the measurement values $z$ obtained by Eve gradually leak information about the secret $s$. Once this point is realized, it becomes evident, particularly for those familiar with information theory, that it is possible to estimate the secret information using the maximum likelihood method. In fact, when using the maximum likelihood method, it is possible to construct the argument without relying on the existence of $r$ or the approximation in Eq.~(\ref{eq:approximation}) assumed in this section. In the next section, we will refine the idea presented here and formulate it in the context of maximum likelihood estimation. By employing tools from information theory, we can provide a more concrete evaluation of the necessary number of measurements.

\subsection{Description via Maximum Likelihood Method}
Let us denote the sequence of measurement values $\{z_1, \cdots, z_N\}$ as $\vec{z}_N$. We consider the conditional probability of obtaining $\vec{z}_N$ given a hypothesized secret information $\hat{s}$:
\begin{equation}\label{eq:probability conditioned by secret information}
p(\vec{z}_N|\hat{s}).
\end{equation}
The maximum likelihood method is a general approach which enables us to estimate the true secret information $s$ by maximizing this likelihood with respect to $\hat{s}$.

Let us examine the specific form of the probability in Eq.~(\ref{eq:probability conditioned by secret information}). Recalling Eq.~(\ref{eq:heterodyne measurement prob}), we introduce the probability distribution:
\begin{eqnarray}\label{eq:each probability conditioned by secret information}
&&\tilde{p}(z_n|\varphi_n(\hat{s})):=\frac{C^{-1}}{|\Upsilon|}\sum_{\upsilon_{\ell}\in \Upsilon} p_{\alpha(\upsilon_\ell,\varphi_n(\hat{s}))}(z_n)\\
&&=\frac{C^{-1}}{|\Upsilon|}\sum_{\upsilon_{\ell}\in \Upsilon}\frac{1}{\pi}\exp\left(-|z_n-\alpha(\upsilon_\ell,\varphi_n(\hat{s}))|^2\right),
\end{eqnarray}
where the sum over $\upsilon_\ell\in \Upsilon$ reflects the fact that Eve does not know the value of the ciphertext $\upsilon_\ell$ and must consider all possibilities equally. Here, $C$ is a normalization constant defined as:
$$
C:=\int_{\mathbf C} dz \frac{1}{|\Upsilon|}\sum_{\upsilon_{\ell}\in \Upsilon}\frac{1}{\pi}\exp\left(-|z-\alpha(\upsilon_\ell,\varphi_n(\hat{s}))|^2\right),
$$
which, due to the condition in Eq.~(\ref{eq:mapping rule for correctness condition}), is a real number close to $1$.
Using Eq.~(\ref{eq:each probability conditioned by secret information}), the probability in Eq.~(\ref{eq:probability conditioned by secret information}) can be expressed as:
$$
p(\vec{z}_N|\hat{s})=\prod_{n=1}^{N}\tilde{p}(z_n|\varphi_n(\hat{s})).
$$
Thus, the negative log-likelihood (NLL) is given by:
\begin{equation}\label{eq:NLL}
\mbox{NLL}(\hat{s}):=-\log p(\vec{z}_N|\hat{s})=-\sum_{n=1}^N \log \tilde{p}(z_n|\varphi_n(\hat{s})).
\end{equation}
Eve can then compute $\mbox{NLL}(\hat{s})$ for each candidate $\hat{s}$ based on the observed measurements $\vec{z}_N$ and choose the $\hat{s}$ that minimizes this score as the estimate of the true secret information $s$.

The theoretical justification for this estimation lies in the fact that the distribution of $\vec{z}_N$ follows $p(\vec{z}_N|\varphi_n(s))$ for the true secret $s$. By taking the expectation of $\mbox{NLL}(\hat{s})$ over $\vec{z}_N$, we have:
\begin{equation}\label{eq:mean score}
\left\langle
\mbox{NLL}(\hat{s})
\right\rangle_{\vec{z}_N}:=-\int_{{\mathbb C}^N} d\vec{z}_N ~p(\vec{z}_N|s) \log p(\vec{z}_N|\hat{s}),
\end{equation}
which is equal to the cross-entropy between the distributions $p(\vec{z}_N|s)$ and $p(\vec{z}_N|\hat{s})$. It is well-known that the cross-entropy is minimized when $p(\vec{z}_N|\hat{s}) = p(\vec{z}_N|s)$, thus justifying the estimation method.

Let us now relate this discussion to the intuitive explanation provided in the previous section. Examining each term $-\log \tilde{p}(z_n|\varphi_n(\hat{s}))$ in Eq.~(\ref{eq:NLL}), we observe the following:
\begin{itemize}
\item If $\varphi_n(\hat{s})\in{\mathcal B}^+(z_n)$, then $\tilde{p}(z_n|\varphi_n(\hat{s}))$ takes a significant value, and thus $-\log \tilde{p}(z_n|\varphi_n(\hat{s}))$ becomes small.
\item If $\varphi_n(\hat{s})\not\in{\mathcal B}^+(z_n)$, then $\tilde{p}(z_n|\varphi_n(\hat{s}))$ becomes extremely small, resulting in a large value for $-\log \tilde{p}(z_n|\varphi_n(\hat{s}))$.
\end{itemize}
Therefore, minimizing Eq.~(\ref{eq:NLL}) is equivalent to finding $\hat{s}$ such that $\varphi_n(\hat{s})\in{\mathcal B}^+(z_n)$ holds for as many $n$ as possible. Unlike the discussion in the previous section, which relied on the existence of $r$ and the approximation in Eq.~(\ref{eq:approximation}), the method presented in this section does not require such assumptions. Hence, the approach described here can be seen as a refinement of the intuitive discussion presented earlier.

To proceed with a more quantitative analysis, let us evaluate Eq.~(\ref{eq:NLL}) in greater detail.
\begin{itemize}
\item First, consider the case where $\hat{s} = s$. For each term in Eq.~(\ref{eq:NLL}), $-\log \tilde{p}(z_n|\varphi_n(s))$, since $\varphi_n(s)$ is expected to be chosen uniformly from ${\mathcal B} = \{b_1, \cdots, b_M\}$, we have:
\begin{itemize}
\item The expected value is:
\begin{equation}\label{eq:a1}
\!\!\!\!\!\!\!\!\!\!\!\!\!\!
a_{s}:=-\frac{1}{M}\sum_{m=1}^M\int_{\mathbb C} dz~\tilde{p}(z|b_m)\log \tilde{p}(z|b_m).
\end{equation}
\item The variance is:
\begin{equation}
\!\!\!\!\!\!\!\!\!\!\!\!\!\!
\Delta^2_{s}\!\!:=\!\!\frac{1}{M}\sum_{m=1}^M \int_{\mathbb C} dz~\tilde{p}(z|b_m)\left[\log \tilde{p}(z|b_m)\right]^2 - a^2_{s}.
\end{equation}
\end{itemize}
Using the central limit theorem, which states that the sum of independent and identically distributed random variables converges to a normal distribution as the sample size increases, the probability density $P_{s}(\lambda)$ of $\mbox{NLL}(s)$ being equal to $\lambda$ for $N \gg 1$ can be approximated as:
\begin{equation}\label{eq:gaussian1}
P^{(N)}_{s}(\lambda)\simeq\frac{1}{\sqrt{2\pi N\Delta^2_{s}}}\exp\left(-\frac{(\lambda-N a_{s})^2}{2N\Delta^2_{s}}\right).
\end{equation}
Note that $a_{s}$, $\Delta_{s}$, and $P^{(N)}_{s}(\lambda)$ do not depend on the specific value of $s$.
\item Next, consider the case where $\hat{s} = s' \neq s$. Similarly, since $\varphi_n(s)$ is chosen uniformly from ${\mathcal B}$ and $\varphi_n(s')$ is also chosen uniformly and independently from ${\mathcal B}$, we have:
\begin{itemize}
\item The expected value is:
\begin{equation}\label{eq:a2}
\!\!\!\!\!\!\!\!\!\!\!\!\!\!
a_{s'}\!\!:=\!\!
-\frac{1}{M}\!\!\sum_{m=1}^M \!\int_{\mathbb C} \!\!dz~\tilde{p}(z|b_m)\frac{1}{M}\!\!\sum_{m'=1}^M\log \tilde{p}(z|b_{m'}).
\end{equation}
\item The variance is:
\begin{equation}\label{eq:delta2}
\!\!\!\!\!\!\!\!\!\!\!\!\!\!
\Delta^2_{s'}\!\!:=\!\!
\frac{1}{M}\!\!\sum_{m=1}^M \!\!\int_{\mathbb C} \!\!\!dz~\tilde{p}(z|b_m)\frac{1}{M}\!\!\!\sum_{m'=1}^{M}\!\!\left[\log \tilde{p}(z|b_{m'})\right]^2\!\! - a^2_{s'}.
\end{equation}
\end{itemize}
Using the central limit theorem, for $N \gg 1$, the probability density $P_{s'}(\lambda)$ of $\mbox{NLL}(s')$ being equal to $\lambda$ can be approximated as:
\begin{equation}\label{eq:gaussian2}
P^{(N)}_{s'}(\lambda)\simeq\frac{1}{\sqrt{2\pi N\Delta^2_{s'}}}\exp\left(-\frac{(\lambda-N a_{s'})^2}{2N\Delta^2_{s'}}\right).
\end{equation}
Again, note that $a_{s'}$, $\Delta_{s'}$, and $P^{(N)}_{s'}(\lambda)$ do not depend on the specific value of $s'$.
\end{itemize}

By applying Jensen's inequality \cite{cover06} to Eq.~(\ref{eq:a2}) and comparing it with Eq.~(\ref{eq:a1}), we immediately obtain:
\begin{equation}
a_{s'} \ge a_{s},
\end{equation}
with the necessary and sufficient condition for equality being:
\begin{equation}\label{eq:condition for equality}
\mbox{"$\tilde{p}(z|b_m)$ are identical and independent of $b_m$."}
\end{equation}

If this condition is satisfied, then $\Delta_{s} \!\!\!=\!\!\! \Delta_{s'}$, and thus $P^{(N)}_{s}(\lambda) = P^{(N)}_{s'}(\lambda)$. In the case of the Y00 protocol, the condition in Eq.~(\ref{eq:condition for equality}) is not satisfied, leading to:
\begin{equation}\label{eq:as as' Y00}
a_{s'}  \gneq a_{s},
\end{equation}
which holds for all variations of the protocol.
Let us evaluate the error probability $P_{err}(N)$ of mistakenly identifying $\mbox{NLL}(s)$ as $\mbox{NLL}(s')$ using Sanov's theorem~\cite{cover06}. Sanov's theorem, a result in large deviation theory, states that the probability of observing an empirical distribution that deviates from the true distribution decreases exponentially with the sample size. Applying Sanov's theorem to the two Gaussian distributions in Eqs. ~(\ref{eq:gaussian1}) and~(\ref{eq:gaussian2}), we obtain:
\begin{equation}
P_{err}(N)\simeq \exp\left[-D\left(P^{(N)}_{s}(\lambda)\|P^{(N)}_{s'}(\lambda)\right)\right],
\end{equation}
where $D(p\|q)$ is the KL-divergence between two probability distributions $p$ and $q$. In our case, this becomes:
\begin{equation}\label{eq:KL divergence}
\!\!\!\!
D\!\left(\!P^{(N)}_{s}\!(\lambda)\|P^{(N)}_{s'}\!(\lambda)\!\right)
\!\!=\!\!\frac{N(a_s-a_{s'})^2\!\!+\!\!\Delta_s^2}{2 \Delta_{s'}^2}\!+\!\log\!\!\left(\!\!\frac{\Delta_{s'}}{\Delta_s}\!\!\right)
\!\!-\!\!\frac{1}{2}.
\end{equation}
For large $N$, the error probability can be approximated as:
\begin{equation}\label{eq:gamma}
P_{err}(N)\simeq \exp\left(-N\Gamma_{ss'}\right),\quad\text{where}\quad
\Gamma_{ss'}:=\frac{(a_s-a_{s'})^2}{2\Delta_{s'}^2}.
\end{equation}
Note that for the Y00 protocol, we have $\Gamma_{ss'} > 0$ due to Eq.~(\ref{eq:as as' Y00}). If the entropy of the secret information $s$ is $\sigma(s)$ nats (approximately $1.44~\sigma(s)$ bits), then to distinguish $s$ uniquely with high probability among $e^{\sigma(s)}$ possible candidates, we require:
$e^{\sigma(s)}P_{err}(N) < 1$, leading to:
\begin{equation}\label{eq:evaluation of Nth}
N\gtrsim N_{th}:=\frac{\sigma(s)}{\Gamma_{ss'}},
\end{equation}
which gives an estimate of the number of measurements needed for successful estimation of the secret information $s$ using the maximum likelihood method.

Let us numerically confirm the effectiveness of the maximum likelihood method through simulations.

\begin{itemize}
\item We assume that the secret information $s$ has been shared between Alice and Bob. The basis sequence is generated as:
$$
\mbox{PRNG}_\Phi(s)=\{\varphi_1(s), \cdots, \varphi_n(s), \cdots, \varphi_N(s)\}.
$$
Each $\varphi_n(s)$ is approximately a random variable uniformly distributed over the set ${\mathcal B}$ in Eq.~(\ref{eq:alphabet of basis}).

\item Since the values of the plaintext $x$ and the encryption key $\kappa$ are not relevant to the vulnerability discussed here, we use a suitable pseudo-random sequence that is approximately uniformly distributed over $\Upsilon$ in Eq.~(\ref{eq:alphabet of ciphertext}):
$$
\{e_1^{(L)}, \cdots, e_n^{(L)}, \cdots, e_N^{(L)}\}.
$$

\item For each pair of $\varphi_n(s)$ and $e_n^{(L)}$, we generate a single sample $z_n \in \mathcal C$ according to the probability distribution:
$$
\frac{1}{\pi} \exp\left(-|z-\alpha(e_n^{(L)},\varphi_n(s))|^2\right).
$$
This process is repeated from $n = 1$ to $N$ to form the sequence of measurement values:
$$
\{z_1, \cdots, z_n, \cdots, z_N\}.
$$
Fig.~\ref{fig_scattering} shows the scatter plots of simulated measurement values for both the P-type and N-type protocols.

\begin{figure}[ht]
\begin{minipage}[b]{0.5\linewidth}
\centering
\includegraphics[width=0.7\linewidth]{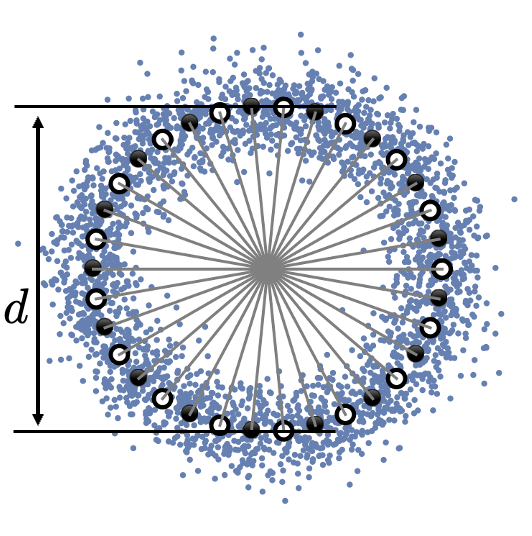}%
\subcaption{P-type ($M=17$, $d=10$)}\label{fig:scatter_a}%
\end{minipage}%
\hfil
\begin{minipage}[b]{0.5\linewidth}
\centering
\includegraphics[width=0.7\linewidth]{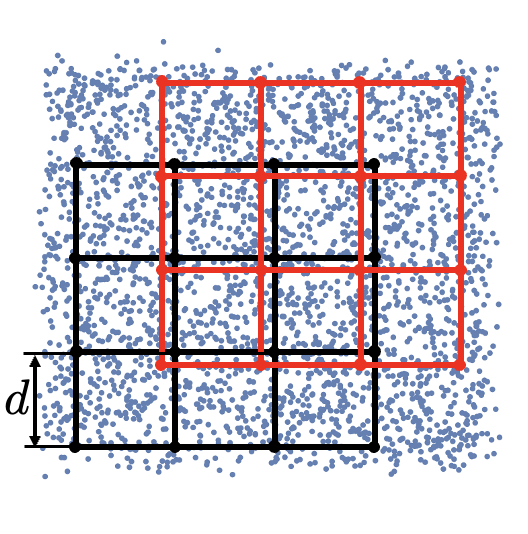}%
\subcaption{N-type ($M=64$, $d=10)$}\label{fig:scatter_b}%
\end{minipage}%
\caption{Scatter plots of simulated data ($N=3000$)}
\label{fig_scattering}
\end{figure}

\item Using the obtained measurement sequence and Eq.~(\ref{eq:NLL}), we compute $\lambda = \mbox{NLL}(s)$ and plot it against $N$. The red plot in Fig.~\ref{fig_nll_n} shows the result.

\item We then use the same measurement sequence to compute $\lambda = \mbox{NLL}(s')$ for a basis sequence generated by a different secret candidate $s'$:
$$
\mbox{PRNG}_\Phi(s')=\{\varphi_1(s'), \cdots, \varphi_n(s'), \cdots, \varphi_N(s')\}.
$$
The gray plots in Fig.~\ref{fig_nll_n} represent the results for 1000 different $s'$ candidates. The blue plot indicates their average.

\item Considering a cross-section of Fig.~\ref{fig_nll_n} for a given $N$, we present the histogram of $\lambda = \mbox{NLL}(s')$ in Fig.~\ref{fig_nll_hist}. The blue and green dashed lines are Gaussian fits based on Eq.~(\ref{eq:gaussian2}), confirming the validity of the central limit theorem. The red arrows indicate the values of $\mbox{NLL}(s)$ for $N=200$ (Fig.~\ref{fig:nll_hist_a}) and $N=800$ (Fig.~\ref{fig:nll_hist_b}), corresponding to the red arrows in Fig.~\ref{fig_nll_n}.

\begin{figure}[ht]
\begin{minipage}[b]{0.5\linewidth}
\centering
\includegraphics[width=0.9\linewidth]{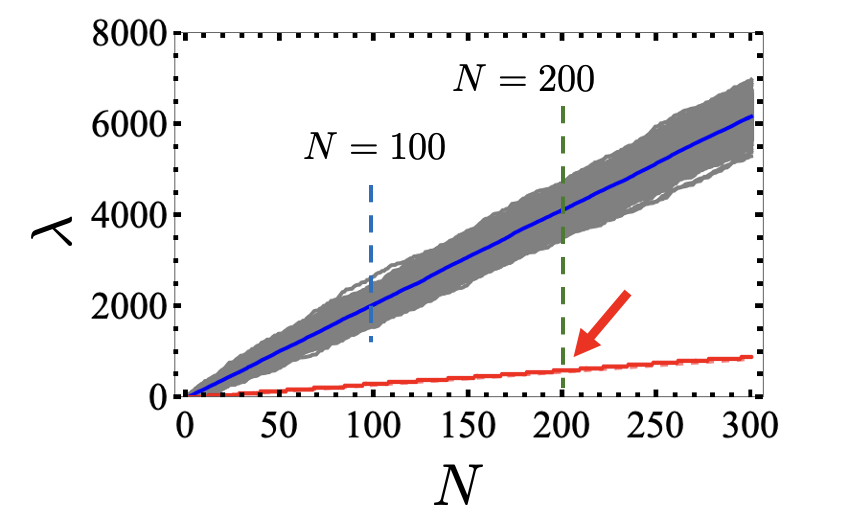}%
\subcaption{P-type ($M=17$, $d=10$)}\label{fig:nll_a}%
\end{minipage}%
\hfil
\begin{minipage}[b]{.5\linewidth}
\centering
\includegraphics[width=0.9\linewidth]{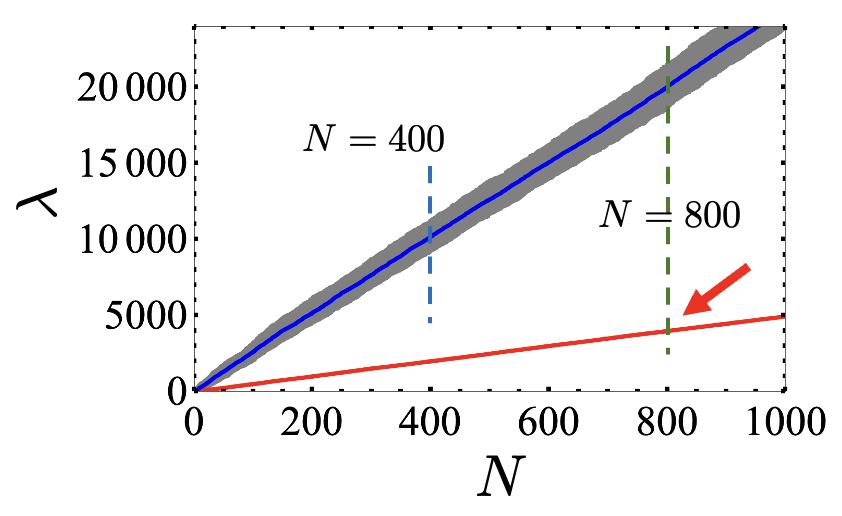}%
\subcaption{N-type ($M=64$, $d=10)$}\label{fig:nll_b}%
\end{minipage}%
\caption{$N$ dependence of $\mbox{NLL}(\hat{s})$ in Eq.(\ref{eq:NLL})}
\label{fig_nll_n}
\end{figure}

\begin{figure}[ht]
\begin{minipage}[b]{0.5\linewidth}
\centering
\includegraphics[width=0.9\linewidth]{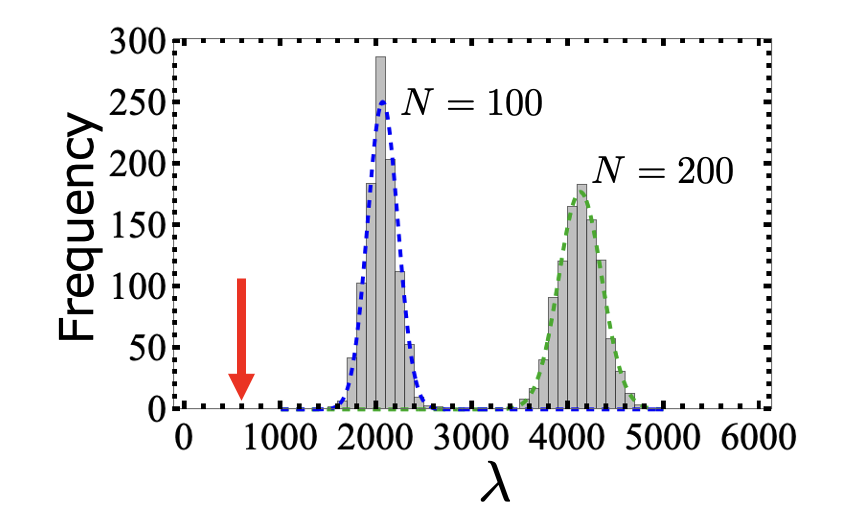}%
\subcaption{P-type ($M=17$, $d=10$)}\label{fig:nll_hist_a}%
\end{minipage}%
\hfil
\begin{minipage}[b]{0.5\linewidth}
\centering
\includegraphics[width=0.9\linewidth]{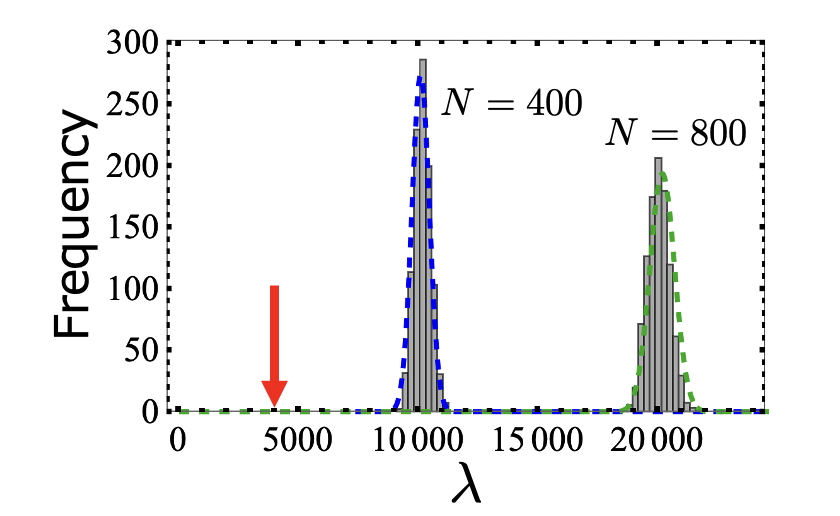}%
\subcaption{N-type ($M=64$, $d=10)$}\label{fig:nll_hist_b}%
\end{minipage}%
\caption{Cross-sections of Fig. \ref{fig_nll_n}}
\label{fig_nll_hist}
\end{figure}

\end{itemize}

From Figs.~\ref{fig_nll_n} and \ref{fig_nll_hist}, it is evident that the value of Eq.~(\ref{eq:NLL}) for the true secret information $s$ is clearly separated from those for any other $s'$, making the estimation of $s$ straightforward. Finally, Table~\ref{table} presents numerical estimates of $a_s$, $a_{s'}$, $\Delta_{s'}$, and $\Gamma_{ss'}$ from Eqs.~(\ref{eq:a1}), (\ref{eq:a2}), (\ref{eq:delta2}), and (\ref{eq:gamma}), respectively, by varying the number of bases $M$ for both P-type and N-type protocols. These values are calculated using the Monte Carlo integration technique \cite{knuth97}, which introduces small statistical fluctuations but provides sufficient precision for the conclusions of this study. Recall that $\Gamma_{ss'}$ determines the required number of measurements for successful estimation of $s$ through Eq.~(\ref{eq:evaluation of Nth}), indicating a relatively small sample size needed for accurate estimation. The table shows that the estimation can be achieved with measurement counts roughly proportional to a few times the entropy $\sigma(s)$. Additionally, as discussed in the previous section, further increasing $M$ beyond a certain value yields diminishing returns in security, as evidenced by the numerical results presented.

\begin{table}[ht]
    \centering
    \begin{tabular}{|c|c|c|c|c|c|}
        \hline
        \multirow{6}{*}{\parbox[c][1.2cm][c]{1.2cm}{P-type\\$d=10$}} 
        & $M$ & $a_{s}$& $a_{s'}$& $\Delta_{s'}$ &$\Gamma_{ss'}$ \\ \cline{1-6}
         & 3 & $2.8\times 10^0$ & $2.0\times 10^1$ & $1.7\times 10^2$& $8.0\times 10^{-1}$\\ \cline{2-6}
         & 9 & $2.8\times 10^0$ & $2.1\times 10^1$ & $2.5\times 10^2$ & $6.3\times 10^{-1}$\\ \cline{2-6}
         & 17 & $2.8\times 10^0$ & $2.1\times 10^1$ & $2.5\times 10^2$ & $6.3\times 10^{-1}$\\ \cline{2-6}
         & 257 & $2.8\times 10^0$ & $2.1\times 10^1$  & $2.5\times 10^2$ & $6.3\times 10^{-1}$ \\ \hline
        \multirow{4}{*}{\parbox[c][1.2cm][c]{1.2cm}{N-type\\$d=10$}} 
        & 16 & $4.9\times 10^0$ & $2.5\times 10^1$ & $2.9\times 10^2$ & $6.8\times 10^{-1}$ \\ \cline{2-6}
         & 64 & $4.9\times 10^0$ & $2.5\times 10^1$ & $3.2\times 10^2$ & $6.3\times 10^{-1}$ \\ \cline{2-6}
         & 256 & $4.9\times 10^0$ & $2.5\times 10^1$ & $3.2\times 10^2$ & $6.2\times 10^{-1}$ \\ \cline{2-6}
         & 1024 & $4.9\times 10^0$ & $2.5\times 10^1$ & $3.3\times 10^2$ & $6.1\times 10^{-1}$ \\ \hline
    \end{tabular}
    \caption{Numerical Estimates\label{table}}
\end{table}

\subsection{Application to the Toy Protocol}
\label{sec:toy implementation}
Let us now consider applying the above attack to the Toy protocol introduced in Sec.~\ref{sec:toy_protocol}. To clearly compare with Y00, we implement the two orthogonal states $\nu_0$ and $\nu_1$ defined in Eq.~(\ref{eq:states for toy}) using two coherent states $|-\alpha\rangle$ and $|+\alpha\rangle$ with a large amplitude $|\alpha|\gg1$. Recall that in the Toy protocol, we had $L = M = 2$ with $\{\upsilon_1 = 0, \upsilon_2 = 1\}$ and $\{b_1 = 0, b_2 = 1\}$. Furthermore, since we set $\rho_{\upsilon_\ell, b_m} = \nu_{\upsilon_\ell \oplus b_m}$, the complex numbers $\alpha(\upsilon_\ell, b_m)$ are defined as:
\begin{equation}
\alpha(0,0)=\alpha(1,1)=-\alpha,\quad
\alpha(0,1)=\alpha(1,0)=+\alpha.
\end{equation}
Thus, we associate the coherent state $|\alpha(\upsilon_\ell, b_m)\rangle$ with each pair $(\upsilon_\ell, b_m)$.
Constructing $\tilde{p}(z|b_m)$ according to Eq.~(\ref{eq:each probability conditioned by secret information}), we have:
\[
\tilde{p}(z|0)=C^{-1}\frac{1}{2}\sum_{\upsilon_{\ell}\in \{0,1\}}\exp\left(-|z-\alpha(\upsilon_\ell,0)|^2\right),
\]
\[
\tilde{p}(z|1)=C^{-1}\frac{1}{2}\sum_{\upsilon_{\ell}\in \{0,1\}}\exp\left(-|z-\alpha(\upsilon_\ell,1)|^2\right).
\]
It is immediately evident that $\tilde{p}(z|0) = \tilde{p}(z|1)$. In other words, since the condition in Eq.~(\ref{eq:condition for equality}) is satisfied, we have $a_s = a_{s'}$ from Eqs.~(\ref{eq:a1}) and (\ref{eq:a2}), and thus $\Gamma_{ss'} = 0$ in Eq.~(\ref{eq:gamma}). This implies that no matter how many measurements are collected, the attack described in the previous section cannot succeed. Hence, the absence of the vulnerability found in Y00 in the Toy protocol indicates that the Toy protocol is more robust than Y00.

It should be noted that conventional stream ciphers, when implemented using coherent states similarly to the Toy protocol, naturally satisfy the condition given
in Eq.~(\ref{eq:condition for equality}).

It should be also noted that limiting the physical states to just two is not essential for implementing the Toy protocol. For instance, we can introduce a $J$-bit binary number $\xi$ as:
\[
\xi := \{\upsilon_{\ell_1} \oplus b_{m_1}, \cdots, \upsilon_{\ell_J} \oplus b_{m_J}\},
\]
and assign a coherent state $|\alpha_{\xi}\rangle$ to each $\xi$, ensuring that $|\alpha_\xi - \alpha_{\xi'}| \gg 1$. It is evident that even when using $2^J$ coherent states, the condition in Eq.~(\ref{eq:condition for equality}) is still satisfied, allowing Alice and Bob to communicate $J$ bits of secret information per coherent state. Although this extension is unrelated to the cryptographic security, it can be implemented using quadrature amplitude modulation (QAM) \cite{thomas74}, similar to what is employed in the N-type Y00.

\section{Conclusion}
\label{sec:conclusion}
In this paper, we highlighted the leakage of secret information from the measurement outcomes in the Y00 protocol, and formulated an attack based on maximum likelihood estimation. This section concludes with discussions on the implications of our findings in the context of security technology.

\begin{itemize}
\item
First, let us examine the practical significance of our findings. The attack we constructed involves an exhaustive search over the space of secret information $\mathcal{S}$ by evaluating Eq.~(\ref{eq:NLL}) for each candidate $s\in\mathcal{S}$. Thus, the computational cost of this attack scales exponentially with the entropy $\sigma(s)$ of the secret information. Depending on the value of $\sigma(s)$, the required computational resources may become prohibitively large, suggesting that this attack may not pose a realistic threat. While this reasoning might appear valid on the surface, it overlooks a critical point.

Primarily, even if this were true, there is no valid justification for adopting a weaker protocol. Since Y00 is demonstrably weaker than the Toy protocol, as shown through our analysis, there is no rational basis for selecting it over the Toy protocol. In \cite{donnet06,ahn07,ahn08}, consistent with the findings of this paper, the possibility of attacks without the use of known plaintext was also initially suggested. These studies, focusing on the original Y00 protocol, demonstrated vulnerabilities such as the feasibility of a fast correlation attack when a linear feedback shift register is used as the PRNG \cite{donnet06}, and information leakage caused by changes in Eve's uncertainty about the bases before and after measurement \cite{ahn07,ahn08}. However, these claims were countered by arguments in \cite{yuen07,nair08} based on the assumption that the computational cost renders the attack impractical. 

If this reasoning were indeed valid, it would imply that the Toy protocol and conventional stream ciphers are not only secure against the attacks discussed in this paper but also against the stronger known-plaintext attacks under the same logic. This would render the Y00 protocol not only redundant but also fundamentally flawed, as it introduces structural vulnerabilities. While the computational cost of exploiting such vulnerabilities may be high under certain conditions, their existence highlights weaknesses in the protocol's design that are absent in its alternatives.

This, in turn, would eliminate any reasonable argument for adopting the weaker Y00 protocol. This argument remains valid even when a secret key sharing protocol such as QKD \cite{gisin02,wolf21} is used alongside Y00 for key refreshment. Even if new secret information is supplied via QKD, there is no reason to waste valuable secret resources on a weaker protocol. In essence, this is a straightforward comparison of relative strengths and weaknesses. Choosing Y00, which inherently consumes more secret information due to its structural flaws, is unjustifiable.

\item
Next, let us evaluate the potential advantages of Y00 stemming from the specific nature of optical communication technology. Y00 was designed from the outset to utilize optical coherent states, requiring Eve to measure using similar optical setups. It might seem that the technical difficulty of such measurements provides a basis for the superiority of Y00. However, it is crucial to separate the difficulty of measurement from the protocol's security. As demonstrated in Sec.~\ref{sec:toy implementation}, the Toy protocol introduced in this paper can also be naturally implemented using optical coherent states. Thus, the high difficulty of Eve's measurement in actual optical communication systems applies equally to both Y00 and the Toy protocol. In such a scenario, opting for a structurally vulnerable protocol like Y00 increases risk, making it more rational to choose structurally secure protocols like the Toy protocol. This discussion shows that Y00 does not possess any inherent advantage due to the specific nature of optical communication, and structurally secure alternatives such as the Toy protocol or conventional stream ciphers are superior choices.

\item
In addition, let us generalize the insights obtained in this paper beyond the Y00 protocol. One of the key results of this paper is the condition expressed in Eq.~(\ref{eq:condition for equality}). We extend this condition to the most general scenario without limiting the specific setup. Consider a situation where Alice encodes a message $x \in {\mathcal X}$ into a quantum state $\mu_{x,s}$ using a secret information $s \in \mathcal S$ and transmits it to Bob. Assume that $x$ follows a uniform distribution over $\mathcal X$. Note that this description encompasses not only the Y00 and Toy protocols but all shared-key cryptographic systems. The necessary and sufficient condition for Eve, who has access only to $\mu_{x,s}$ without knowledge of $x$, to gain no information about $s$ from any quantum measurement is:
\begin{equation}
\text{"$\mu(s)$ are identical and independent of $s$"}
\end{equation}
where
\[
\mu(s) := \frac{1}{|{\mathcal X}|}\sum_{x \in {\mathcal X}} \mu_{x,s}.
\]
This condition is a generalization of the condition in Eq.~(\ref{eq:condition for equality}). In existing cryptographic systems, this condition has been so trivially satisfied that it has received little attention. However, if this condition is not met, then, as discussed in this paper, attacks of the type we have analyzed could exist at least in principle. This would result in a weaker scheme that fails to achieve the level of security traditionally provided by conventional stream ciphers. This condition will be useful in evaluating new shared-key cryptographic systems employing physical states like Y00, irrespective of the specific details of their implementation.

\item
Lastly, we note the critical importance of evaluating individual cryptographic schemes, even in highly controlled environments such as national defense. Operational secrecy, which involves concealing the inner workings of cryptographic systems, and the diversity of cryptographic methods, which uses multiple distinct schemes to mitigate systemic risks, are valuable strategies that add layers of security. However, they cannot fully compensate for unassessed vulnerabilities in individual schemes. Such vulnerabilities, if exploited, can compromise the security of the broader system. 

Furthermore, prior analyses of the Y00 protocol have often relied on assumptions that inadequately address its security. For instance, claims that the protocol achieves a strong "dual structure" through quantum mechanical properties or that it ensures resistance to quantum computational attacks have been presented without rigorous verification. 
These assertions, while misguidedly appealing, lack the empirical or analytical support necessary to substantiate their validity. Overestimating the security of a protocol based on such incomplete or superficial evaluations not only fosters misplaced trust but also risks exposing systems to avoidable vulnerabilities.

This highlights the importance of rigorous and comprehensive evaluations that do not merely aim to justify existing designs but critically examine their limitations and vulnerabilities. Such evaluations are particularly crucial for cryptographic schemes that claim innovative features, as these features must be scrutinized to ensure they do not introduce unforeseen weaknesses.
Rigorous evaluation ensures that these vulnerabilities are identified and mitigated before deployment. Moreover, such evaluation advances the field of cryptography by uncovering design principles that strengthen future systems, ensuring that security is built on a foundation of sound scientific analysis.
\end{itemize}

This paper was motivated by the proposal of the Y00 protocol and the subsequent series of research developments. As stated at the beginning of this paper, scientific progress relies on a cycle of proposals and verification. In this work, we presented a new perspective on the structural vulnerabilities of Y00, aiming to contribute to the healthy evolution of information security technology through rigorous examination. We hope that this study contributes, even in a small way, to the realization of a secure and trustworthy information society based on sound scientific discussions.

\section*{Acknowledgements}
The author would like to express his heartfelt gratitude to Professor Tsutomu Matsumoto for sharing valuable insights that inspired this research, for providing the motivation to report its findings, and for his meticulous review of the manuscript, which greatly improved its clarity. Sincere thanks are also extended to Dr. Shinichi Kawamura for his thorough review of the manuscript and his invaluable advice on its structure. Additionally, the author acknowledges OpenAI's ChatGPT-4 for its assistance in refining the tone and clarity of the English expressions in this manuscript. While the content and scientific analysis remain the sole responsibility of the author, the feedback provided by the tool was instrumental in enhancing the overall readability of the paper.

\bibliographystyle{unsrt}

\end{document}